\documentclass{pasj00}

\begin{document}
\SetRunningHead{Asai et al.}{Spectral state transition in NS-LMXBs}
\Received{2015/04/19}
\Accepted{2015/06/05}

\title{X-ray variability with spectral state transitions in NS-LMXBs observed with MAXI/GSC and Swift/BAT}

\author{
Kazumi \textsc{Asai},\altaffilmark{1}
Tatehiro \textsc{Mihara},\altaffilmark{1}
Masaru \textsc{Matsuoka},\altaffilmark{1}
and
Mutsumi \textsc{Sugizaki},\altaffilmark{1}
}
\altaffiltext{1}{MAXI team, RIKEN, 2-1 Hirosawa, Wako, Saitama 351-0198}
\email{kazumi@crab.riken.jp}
%

\maketitle

\begin{abstract}

X-ray variabilities with spectral state transitions
in bright low mass X-ray binaries containing a neutron star
are investigated 
by using 
the one-day bin light curves of
MAXI/GSC (Gas Slit Camera)
and Swift/BAT (Burst Alert Telescope).
Four sources (4U~1636$-$536, 4U~1705$-$44, 4U~1608$-$52, and GS~1826$-$238)
exhibited small-amplitude X-ray variabilities with spectral state transitions.
Such ``mini-outbursts'' were characterized by smaller amplitudes
(several times) and shorter duration (less than several tens of days)
than those of ``normal outbursts.''
Theoretical model of disk instability by Mineshige and Osaki 
(PASJ, 37, 1, 1985) predicts both large-amplitude
outbursts and small-amplitude variabilities.  
We interpret the normal outbursts as the former prediction of this model,
and the mini-outbursts as the latter.
Here, we can also call the mini-outburst as
``purr-type outburst'' referring to theoretical work.
We suggest that similar variabilities lasting for 
several tens of days without spectral state transitions,
which are often observed in the hard state,
may be a repeat of mini-outbursts.
\end{abstract}

\section{Introduction}

The X-ray spectra of low mass X-ray binaries with a neutron star (NS-LMXBs)
mainly depend on the X-ray luminosities (e.g., \cite{Lewin1997}).
High luminosity ($\gtrsim 10^{37}$~erg s$^{-1}$)
dominated by thermal component is called ``soft state,''
whereas low luminosity ($\lesssim 5\times 10^{36}$~erg s$^{-1}$)
dominated by the Comptonized component is called ``hard state.''
In general, 
the luminosity is thought to reflect the mass-accretion rate and
then the spectral state (soft or hard) corresponds to a
state of the inner accretion disk (optically thick or thin)
(e.g., \cite{Mitsuda1989}; \cite{White1988};
\cite{Done2007}; \cite{Matsuoka2013}).
Spectral states are a key to determine
the state of the inner accretion disk.

Bright NS-LMXBs have been classified into two groups,
Z sources and Atoll sources, based on their behavior on the color--color
diagram and hardness--intensity diagram \citep{Hasinger1989}.
Z sources are generally bright and sometimes become close to
the Eddington luminosity ($L_{\rm E}$).
On the other hand, Atoll sources are generally less bright,
and some of them exhibit spectral state transitions (soft or hard).
In the color--color diagram,
the shape can be divided into two main regions, ``banana'' and ``island.''
The spectrum is usually softer in the banana than in the island.

Spectral state transitions
have been studied for transient X-ray
binaries including a neutron star or a black hole
(e.g., \cite{Maccarone2003a}; \cite{Maccarone2003b};
\cite{Yu2009}; \cite{Tang2011}; \cite{Asai2012}). 
Usually, these transitions are observed
in the rise and decay phases of outbursts, 
when the luminosity changes by 2--5 orders of magnitude
during outbursts.
\citet{Maccarone2003a} reported that the soft-to-hard transition in
the decay phase of the outburst occurs at 
1\%--4\% of $L_{\rm E}$.
For the hard-to-soft transition in the rise phase of the outburst,
the transition luminosity tends to be larger than that
in the soft-to-hard transition
(\cite{Miyamoto1995}; \cite{Maccarone2003b}; \cite{Asai2012}).
The spectral state transitions have been also observed in 
four persistent NS-LMXBs: 
4U~1636$-$536 \citep{Shih2005},
GS~1826$-$238 \citep{Nakahira2014},
4U~1705$-$44 \citep{Lin2010},
and 4U~1820$-$30 (\cite{Wen}; \cite{Titarchuk}).
The X-ray variability with spectral state transitions
means the change of the intrinsic mass accretion rate.

Recurrent outburst behavior is interpreted by the 
thermal viscous disk instability model (see \cite{Lasota2001} for a review).
In this model, local disk instabilities (S-shaped curves
in the surface density and mass accretion rate diagram)
are caused by the partial ionization of hydrogen and helium.
The local instabilities propagate over the whole disk,
and when it reaches the innermost part of the accretion disk,
an outburst occurs. 
Mineshige and Osaki (1983, 1985) investigated
the disk instability model for outbursts of dwarf novae.
They reported that instabilities on a constant-$\alpha$ disk
yield only small-amplitude variabilities,
where $\alpha$ is the standard viscosity parameter.
This was also pointed out by \citet{Smak1984} and \citet{Meyer1984}.
On the other hand, by assigning high $\alpha$ in a hot state
(upper branch in S-shaped curve)
and low $\alpha$ in a cool state (lower branch),
the width of the unstable branch (middle branch) is extended and
the instability can propagate through the whole disk.
The two-$\alpha$ disk leads to the large-amplitude outbursts.
Mineshige and Shields (1990a) studies the similar variability
for the disk instability in AGN, and then, they named
the small and the large amplitude variabilities as ``purr type'' and
``roar type,'' respectively.
In this paper, we propose that some small amplitude variabilities
(mini-outbursts) detected in NS-LMXBs are considered to be
the purr-type outburst. 

Smak (1982, 1983) showed that
instability occurs below a critical effective
temperature depending on the mass accretion rate.
They classified the cataclysmic variables into two objects
on the outer disk radius versus mass accretion rate diagram: 
stationary-accretion objects (novae and nova-like stars)
and non stationary-accretion objects (dwarf novae). 
External heating by central object may be ignored
in white dwarfs, but is non-negligible in NS-LMXBs
because the disk is strongly irradiated.
In other words, 
instability is determined by not only the mass accretion rate,
but also the X-ray luminosity.
Tuchman, Mineshige, and Wheeler (1990) calculated
the instability in consideration
of the effect of irradiation, and
generated the S-shaped curves at different irradiation temperatures.
They concluded that 
for sufficiently strong irradiation (irradiation temperature
above 10000~K), the hydrogen and helium atoms are completely ionized,
thus, the thermal instability is suppressed.
\citet{Paradijs1996} also calculated the effect of external heating
and separated persistent NS-LMXBs from transient NS-LMXBs 
on the orbital period versus X-ray luminosity diagram.

The present study focuses
on the properties of the spectral state transitions of  
transient and persistent NS-LMXBs.
The analyzed data are 
MAXI \citep{Matsuoka2009}/GSC
(Gas Slit Camera: \cite{Mihara2011}; \cite{Sugizaki2011})
\footnote{$<$http://maxi.riken.jp/$>$.}
and Swift \citep{Gehrels2004}/BAT
(Burst Alert Telescope: \cite{Barthelmy2005})
\footnote{$<$http://heasarc.gsfc.nasa.gov/docs/swift/results/transients/$>$.}
from 2009 August 15 (MJD = 55058) to 2014 December 31 (MJD = 57022).
Section~2 describes the data analysis and the properties of
spectral state transition.
In section~3,
we discuss the properties of the transition
based on the disk instability model.
Conclusion is presented in section~4. 

\section{Analysis and results}

\subsection{Source selection}

\begin{table*}
\caption{Properties of the seven NS-LMXBs analyzed in this paper.}
\label{tab1}
\begin{center}
\begin{tabular}{lcccc}
\hline
Name & State & P$_{\rm orbital}$ (hr) & Distance (kpc) & Reference\footnotemark[$*$] \\
\hline
4U~1636$-$536 & Persistent   & 3.80      & 6    & (1) \\
4U~1705$-$44  & Persistent   & unknown   & 7.4  & (1) \\
4U~1608$-$52  & Transient    & 12.89     & 4.1  & (1), (2) \\
GS~1826$-$238 & Persistent   & 2.088     & 7    & (1), (3) \\
Aql~X-1       & Transient    & 18.95     & 5    & (1) \\ 
XTE~J1709$-$267    & Transient & unknown & 8.8  & (1) \\
4U~1820$-$30  & Persistent   & 0.19      & 7.6  & (1) \\
\hline
\multicolumn{5}{@{}l@{}}{\hbox to 0pt{\parbox{130mm}{\footnotesize
\par\noindent
\footnotemark[$*$]
(1) \cite{Liu2007}, (2) \cite{Galloway2008}, and (3) \cite{Cocchi2011}. 
}\hss}}
\end{tabular}
\end{center}
\end{table*}

\begin{table*}
\caption{Summary of spectral states for seven NS-LMXBs.}
\label{tab2}
\begin{center}
\begin{tabular}{lcccc}
\hline
Name & Thresholds &
\multicolumn{2}{c}{
Average $L_{\rm 2-10~keV}$
\footnotemark[$\dagger$]}
& Outburst \\
     & of hardness ratio\footnotemark[$*$] & Hard state & Soft state
& type\footnotemark[$\ddagger$] \\
      \hline
4U~1636$-$536 & 0.253 & $2.94\pm0.02$ &  $6.95\pm0.02$ & mini \\ 
4U~1705$-$44\footnotemark[$\S$] & 0.355 & $4.22\pm0.04$ & $27.77\pm0.04$ & normal \\
              &       &     & $7.22\pm0.06$ & mini \\ 
4U~1608$-$52\footnotemark[$\S$] & 0.224 & $1.07\pm0.01$ & $11.43\pm0.03$ & normal \\
             &       &     & $2.25\pm0.05$ & mini \\ 
GS~1826$-$238 & 0.2   & $2.95\pm0.01$ &  $8.15\pm0.16$ & mini \\
Aql~X-1       & 0.222 & $1.55\pm0.02$ & $23.6\pm0.07$ & normal \\   
4U~1820$-$30  & 0.2   & $17.61\pm0.02$ & $34.15\pm0.03$ & other \\
XTE~J1709$-$267 & 0.282 & $<$ 1.12 & $12.5\pm0.1$ & normal \\ 
\hline
\multicolumn{4}{@{}l@{}}{\hbox to 0pt{\parbox{120mm}{\footnotesize
\par\noindent
\footnotemark[$*$]
Threshold between soft and hard states, which was determined
using the distribution of the BAT/GSC hardness ratio in figure~\ref{fig1}.  
\par\noindent
\footnotemark[$\dagger$]
Average luminosity in the 2--10~keV band in the unit of $10^{36}$~erg s$^{-1}$.
The luminosity of hard and soft states are the averages of
the bins in the left and right part of the vertical dotted line
in figure~\ref{fig1}, respectively.
\par\noindent
\footnotemark[$\ddagger$]
``normal'' denotes that average luminosity of soft state is above $\sim 10^{37}$ erg s$^{-1}$
and ``mini'' denotes that that is below $\sim 10^{37}$ erg s$^{-1}$.
For 4U~1820$-$30, we note ``other'' because average luminosity
in both soft and hard state are above $\sim 10^{37}$ erg s$^{-1}$.
\par\noindent
\footnotemark[$\S$]
Those sources had two types of soft states with clearly different luminosities.
}\hss}}
\end{tabular}
\end{center}
\end{table*}

\begin{figure*}
 \begin{center}
\includegraphics[width=5cm]{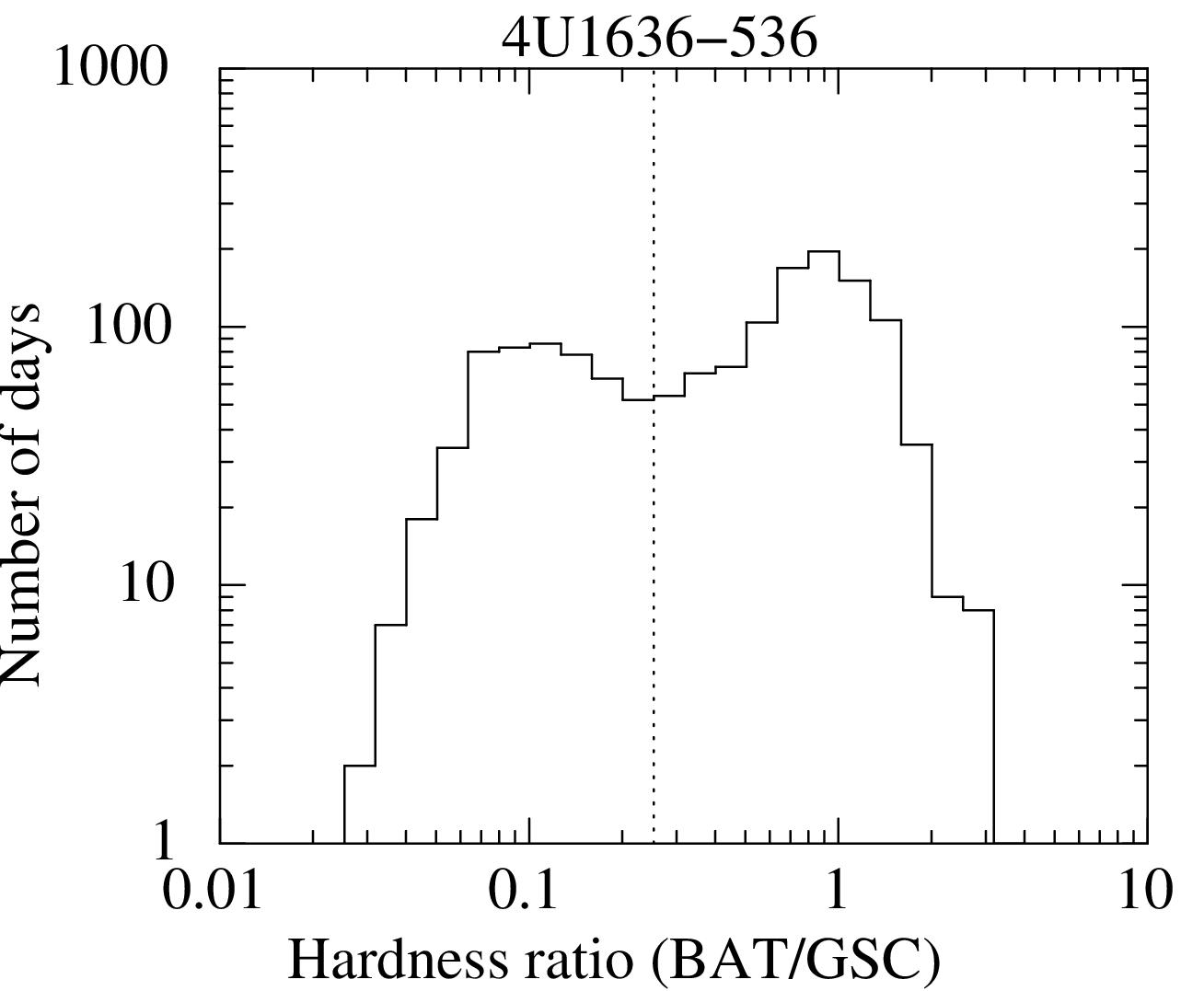} 
\includegraphics[width=5cm]{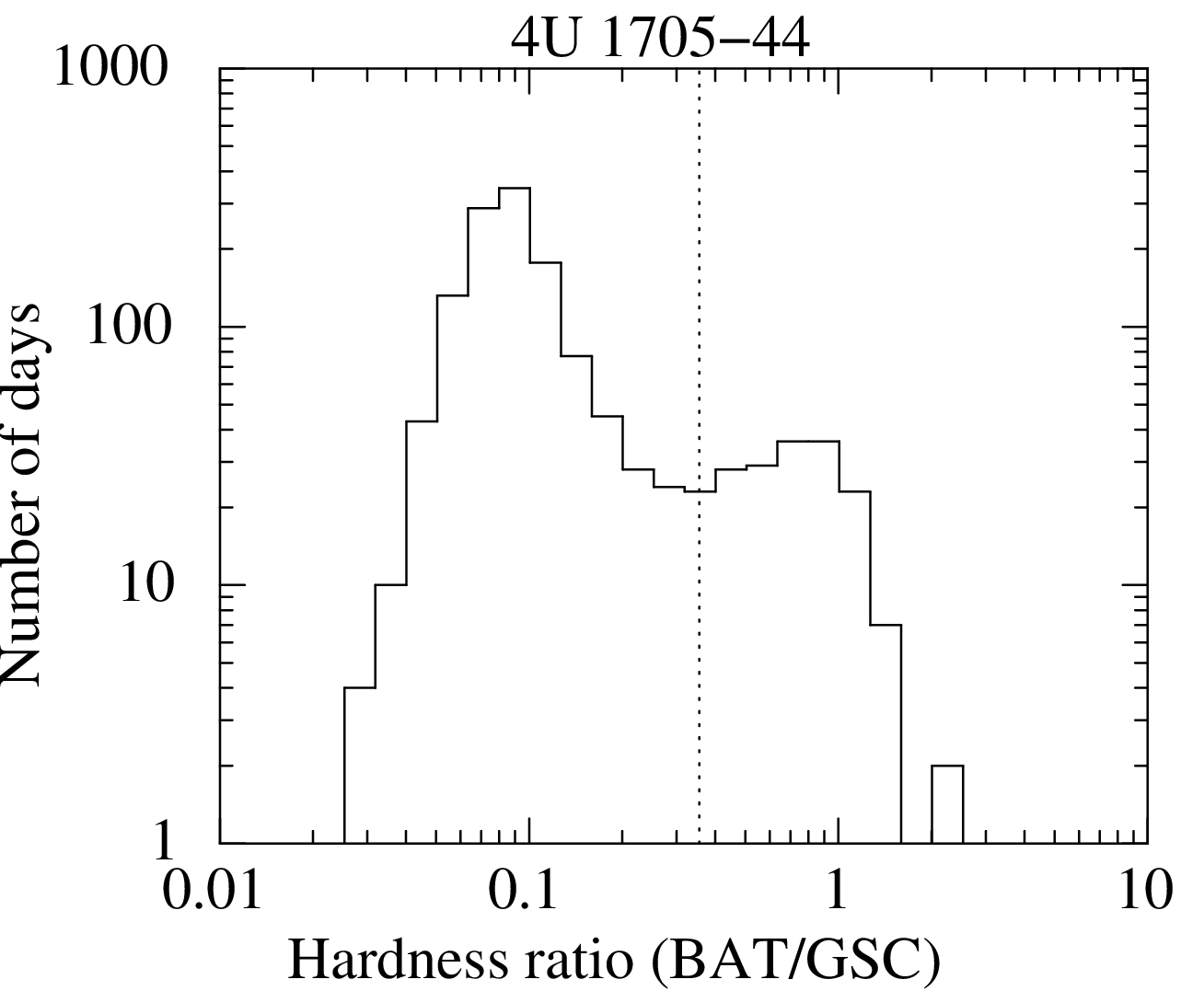}
\includegraphics[width=5cm]{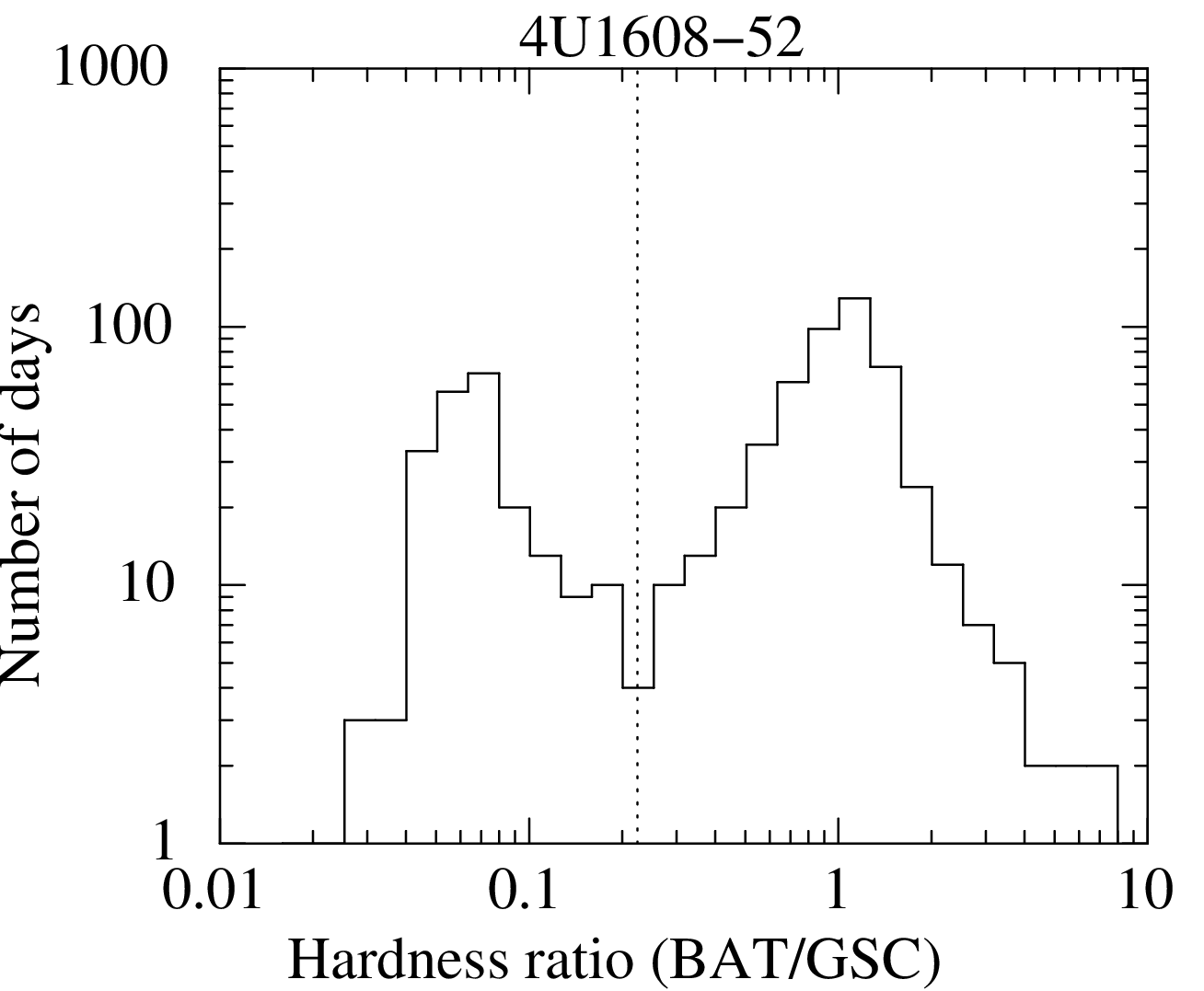} 
\includegraphics[width=5cm]{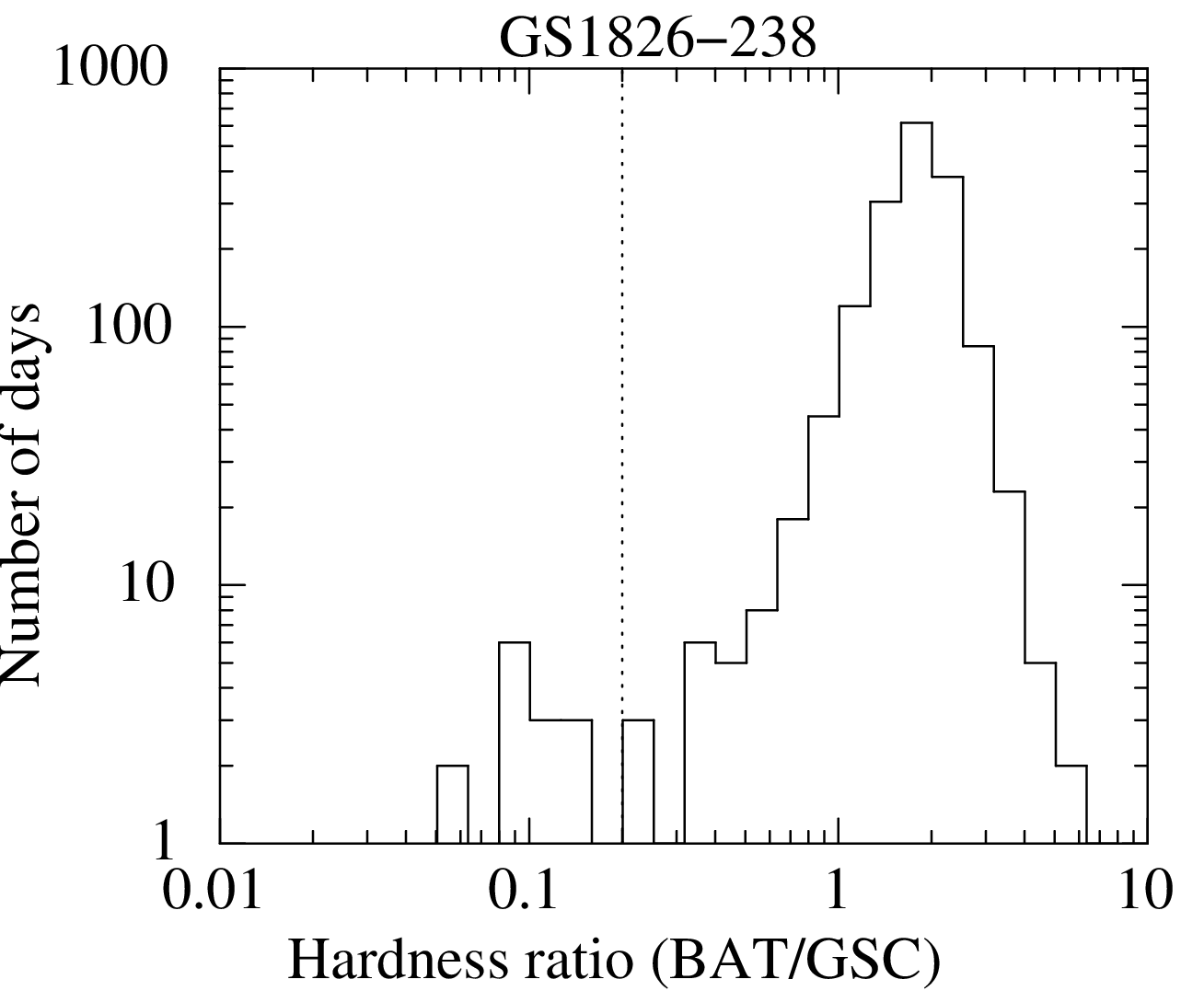} 
\includegraphics[width=5cm]{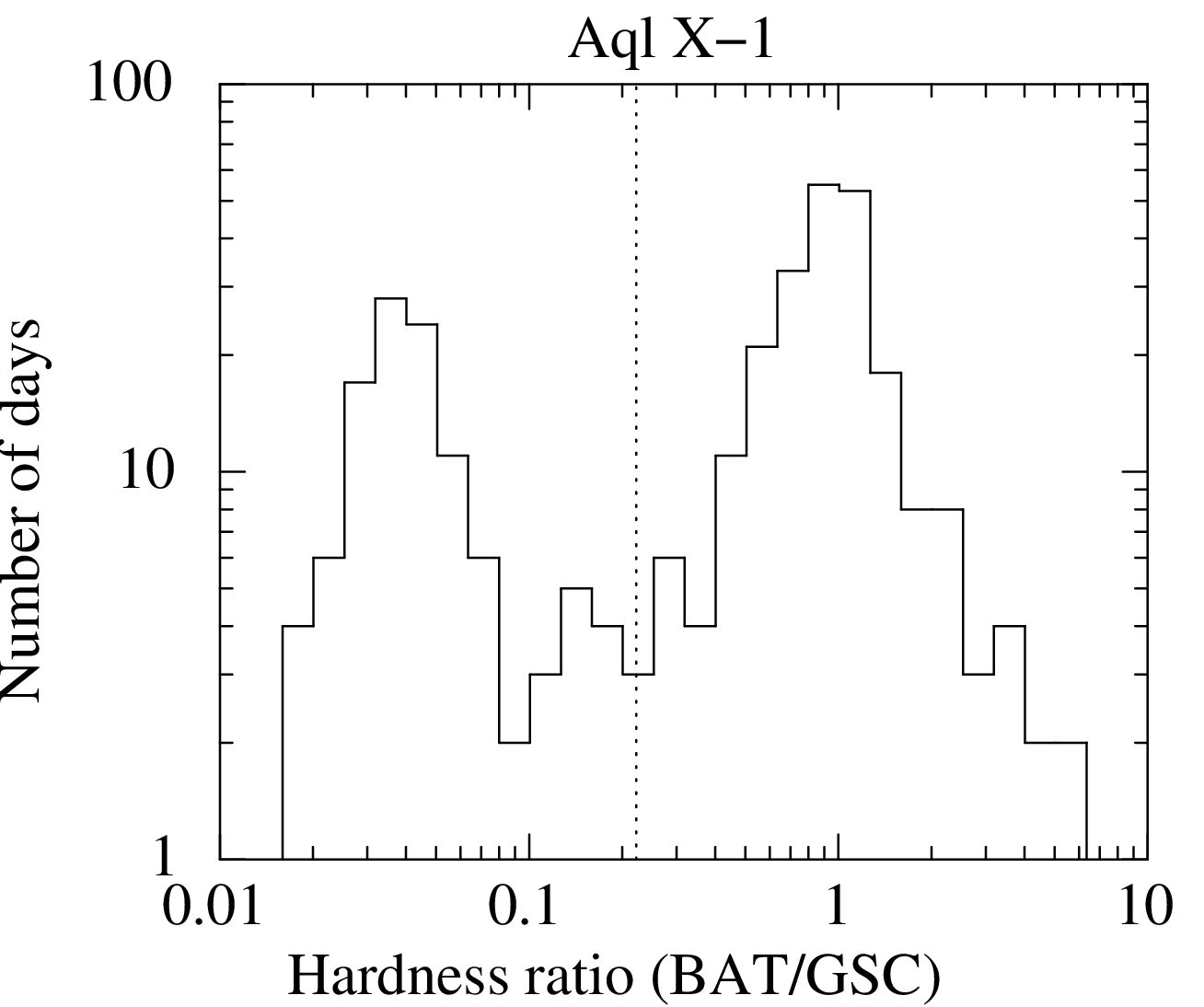} 
\includegraphics[width=5cm]{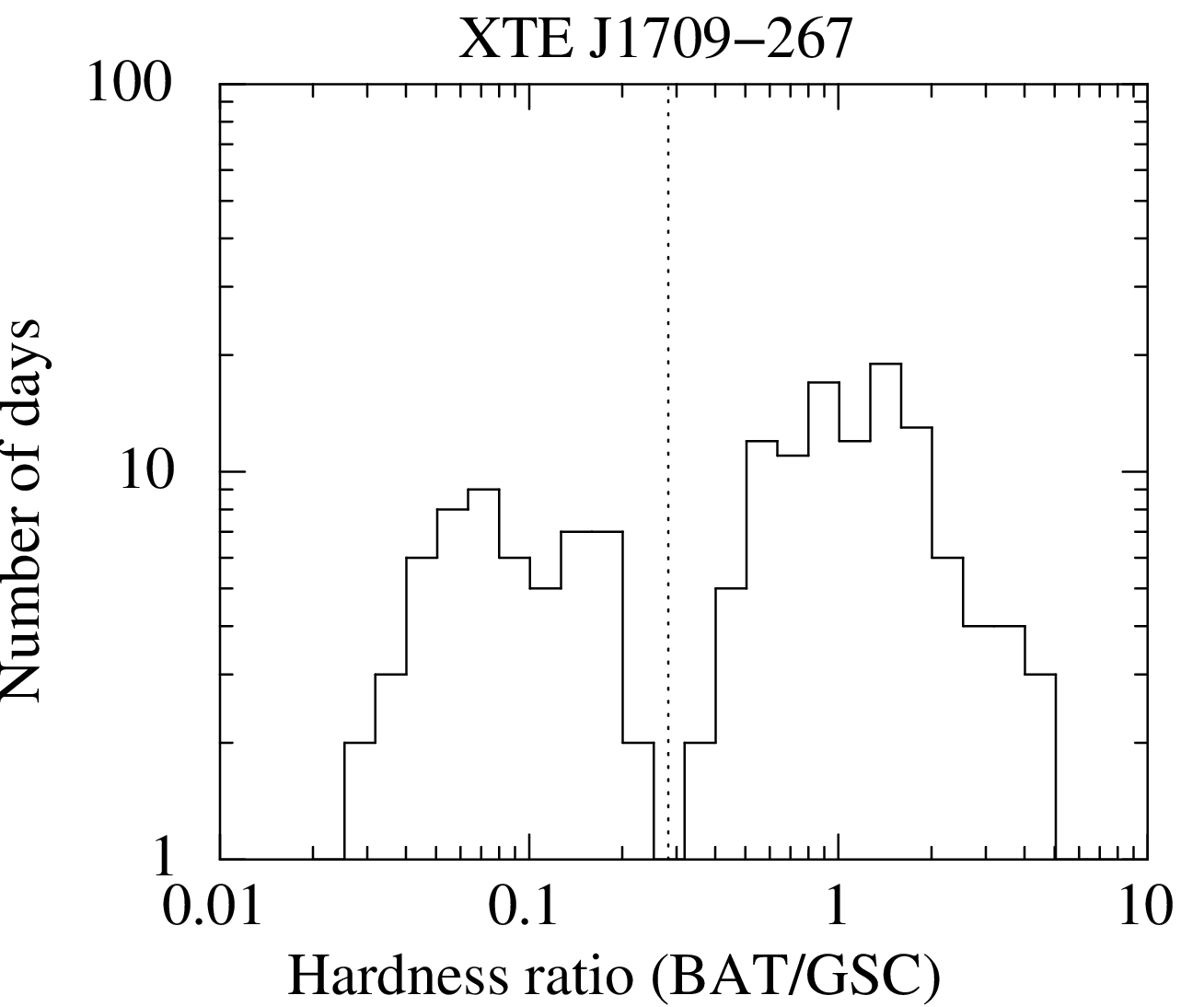} 
\includegraphics[width=5cm]{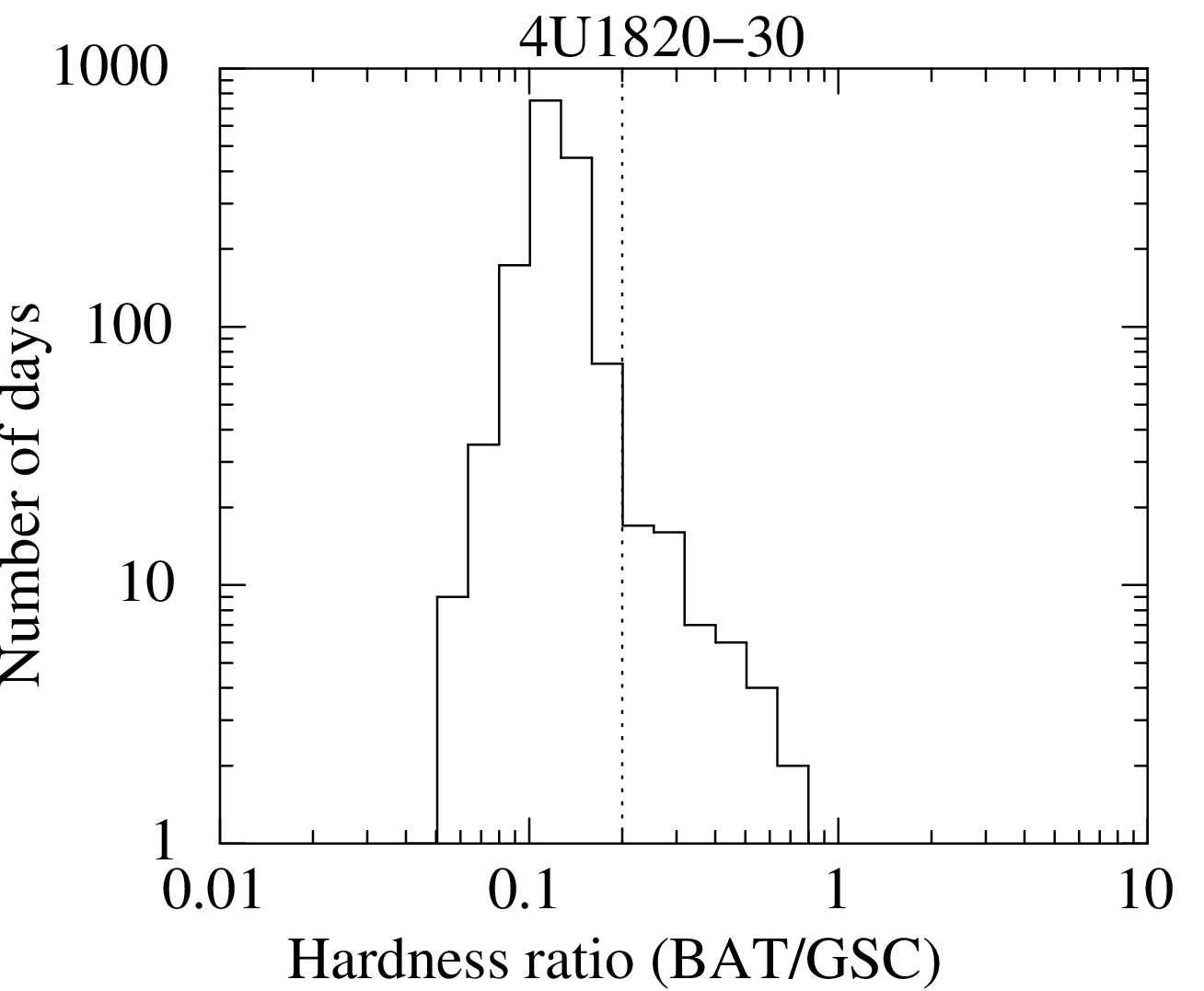} 
 \end{center}
\caption{Distributions of hardness ratios of BAT/GSC for seven NS-LMXBs.
Distributions were constructed from data 
with a significance $>$ 1~$\sigma$.
The vertical dotted lines represented the threshold between
soft and hard states (see text for explanation).
}
\label{fig1}
\end{figure*}


We investigated spectral state transitions in bright NS-LMXBs
using the one-day bin light curves of MAXI/GSC and Swift/BAT.
The hardness ratio of BAT (15--50~keV) and GSC (2--10~keV) is suitable
for identifying spectral state transitions in NS-LMXBs, as 
was employed by Yu and Yan (2000), Tang, Yu, and Yan (2011), and
\citet{Asai2012}.
The obtained count rates of GSC and BAT were converted to luminosities
by assuming a Crab-like spectrum \citep{Kirsch2005}
and the distances listed in the LMXB catalogue \citep{Liu2007}.
\footnote{For 4U~1608$-$52, two values (3.6 and 4.1~kpc) were listed.
In this paper, we employ 4.1~kpc according to \citet{Galloway2008}.
For GS~1826$-$238, the value of 4-8~kpc was listed, but we employ
7~kpc \citep{Cocchi2011}.
}
This assumption is acceptable in the hard state,
because the energy spectrum is dominated by the Comptonized emission
approximated by a power-law.
On the other hand, in the soft state,
the energy spectrum is dominated by the thermal emission.
The obtained luminosity by assuming Crab-like spectrum
is underestimated in the 2--10~keV band and
is overestimated in the 15--50~keV band.
When the spectrum in the soft state is a blackbody emission
with $kT \sim $2--2.5~keV,
the obtained luminosity in the 2--10~keV band is 0.7 of that of
the blackbody emission. 
In the 15--50~keV, the obtained luminosity
is 1.4 of that of the blackbody emission.
As a result, the hardness ratio of two energy bands is overestimated
by 2.0 times.
Moreover, we need a bolometric correction to discuss the luminosities.
The correction factor from a 2--10~keV luminosity to
a bolometric (0.1--100~keV)
one is 1.4--1.7 for the blackbody spectrum assumed above.
In this paper, those corrections should be considered to discuss
the real luminosity.

\begin{figure*}
 \begin{center}
\includegraphics[width=16cm]{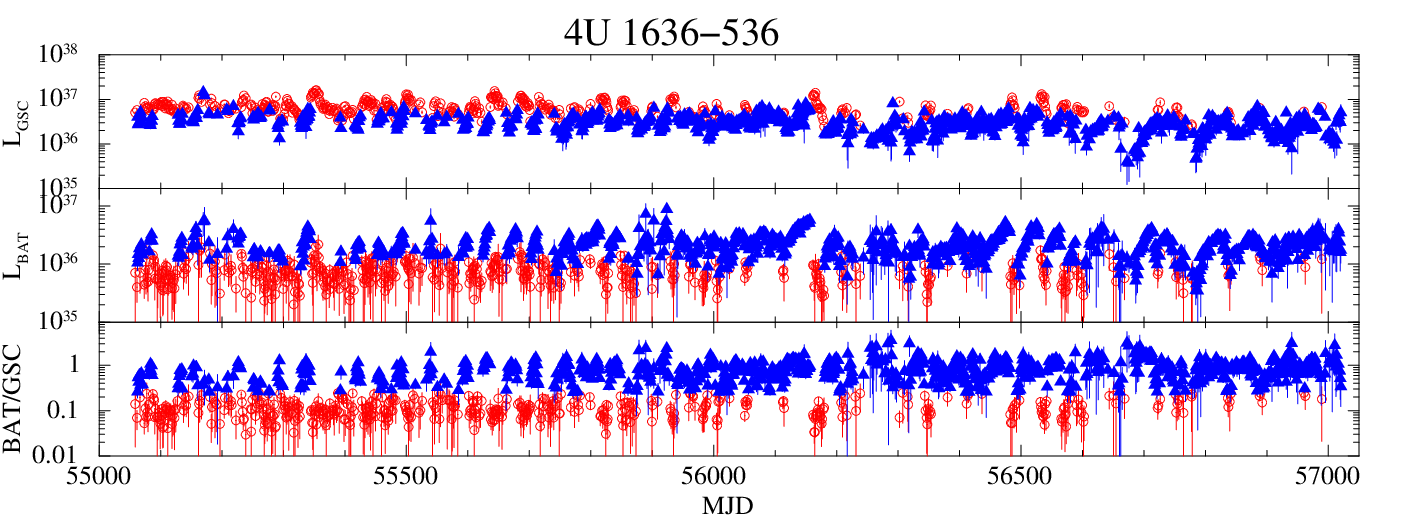}  
\includegraphics[width=16cm]{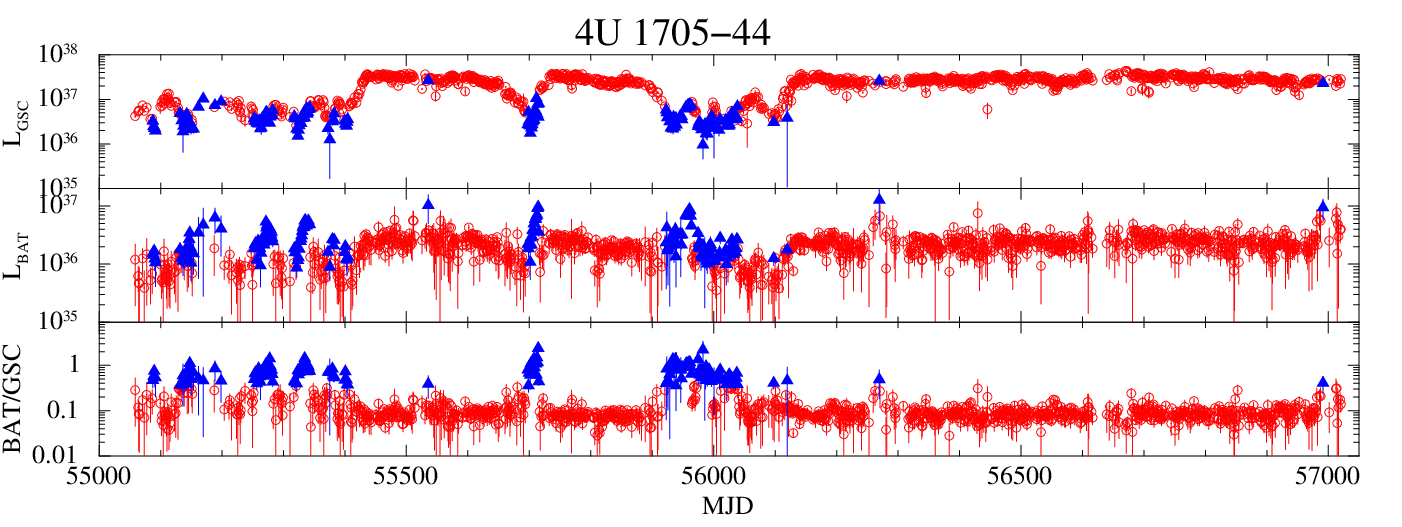} 
\includegraphics[width=16cm]{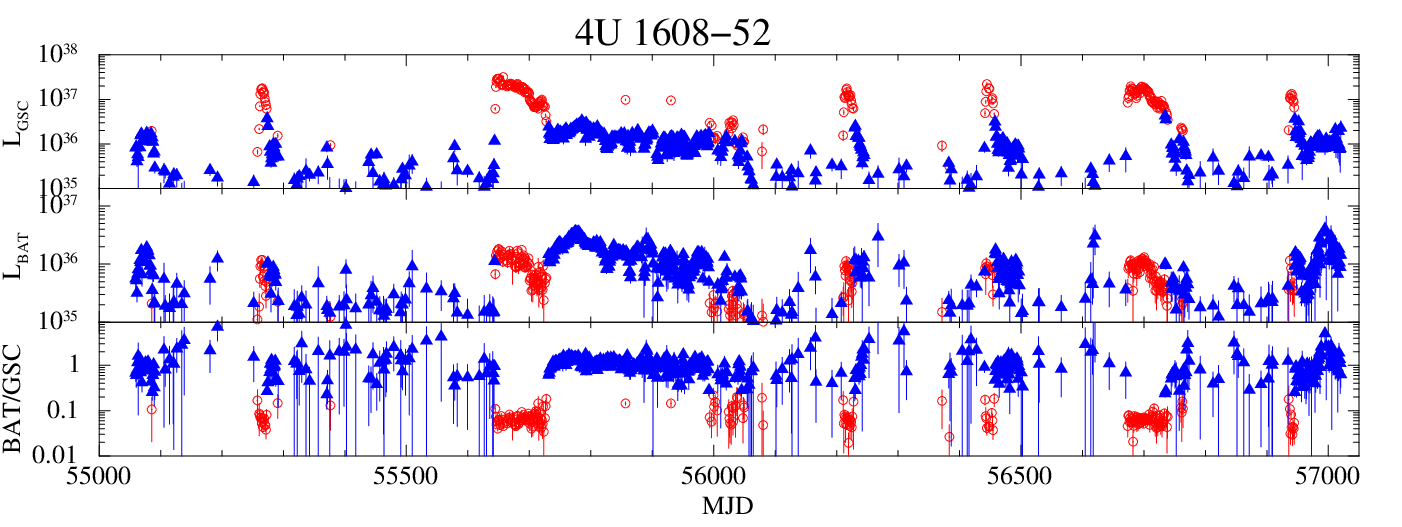}  
\includegraphics[width=16cm]{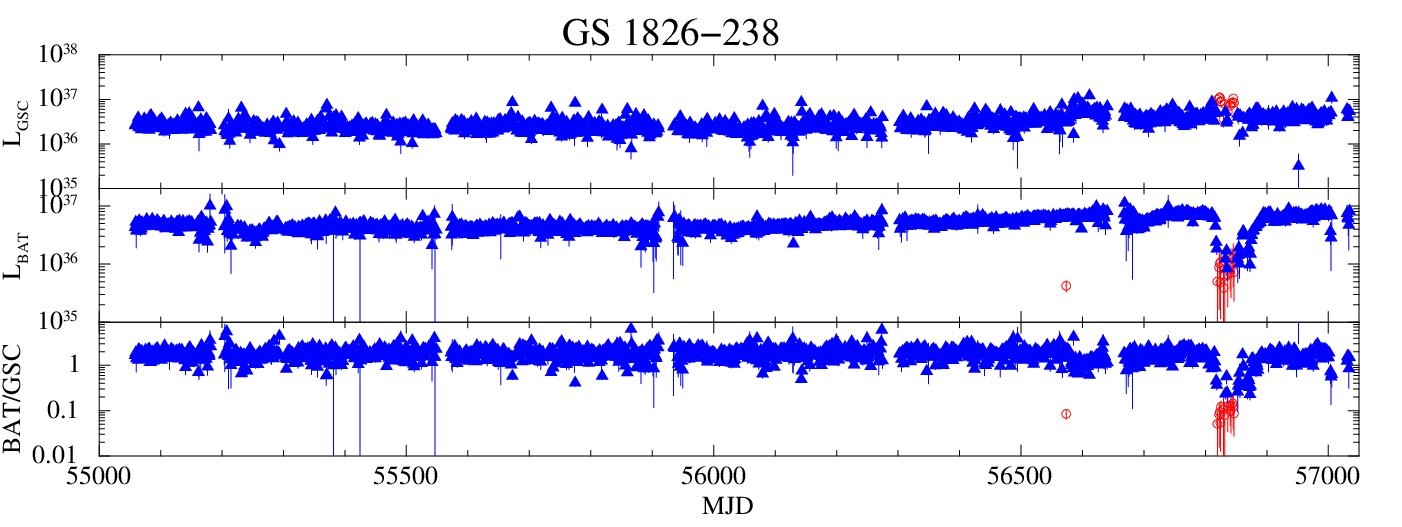}  
 \end{center}
\caption{One-day GSC light curves in the 2--10~keV band,
one-day BAT light curves in the 15--50~keV band,
and the hardness ratio (BAT/GSC)
of 4U~1636$-$536, 4U~1705$-$44, 4U~1608$-$52, and GS~1826$-$238.
$L_{\rm GSC}$ and $L_{\rm BAT}$ are the luminosities in units of erg s$^{-1}$.
Vertical error bars represent 1-$\sigma$ statistical uncertainty. 
Red circles and blue triangles represent data of soft and hard states,
respectively. (Color online)}
\label{fig2}
\end{figure*}

\begin{figure*}
 \begin{center}
\includegraphics[width=16cm]{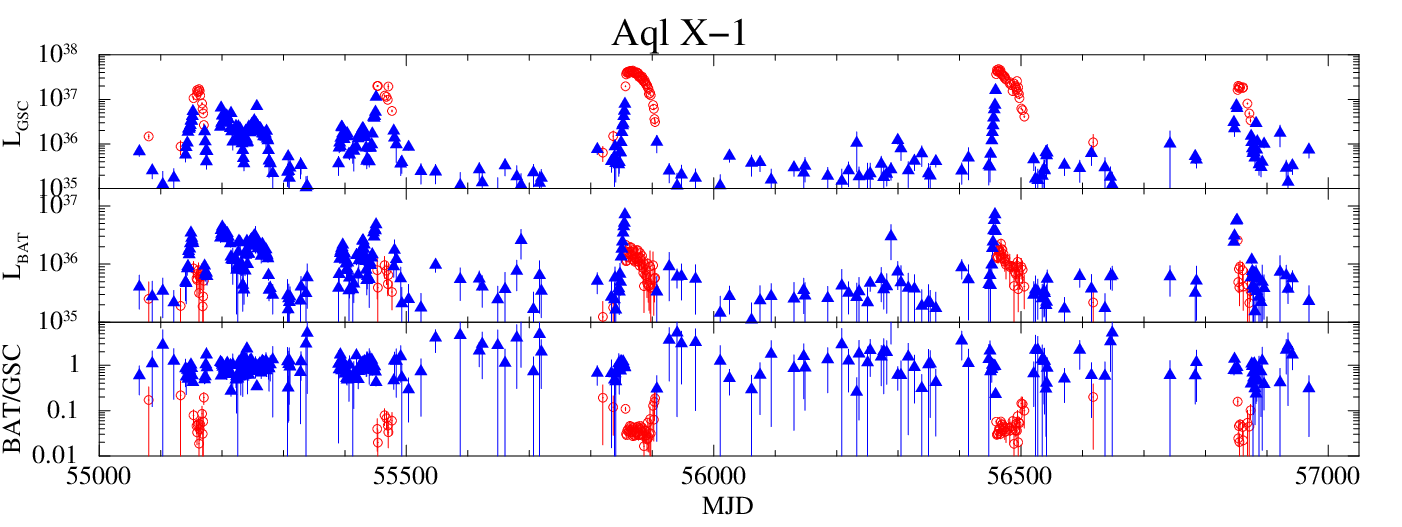}  
\includegraphics[width=16cm]{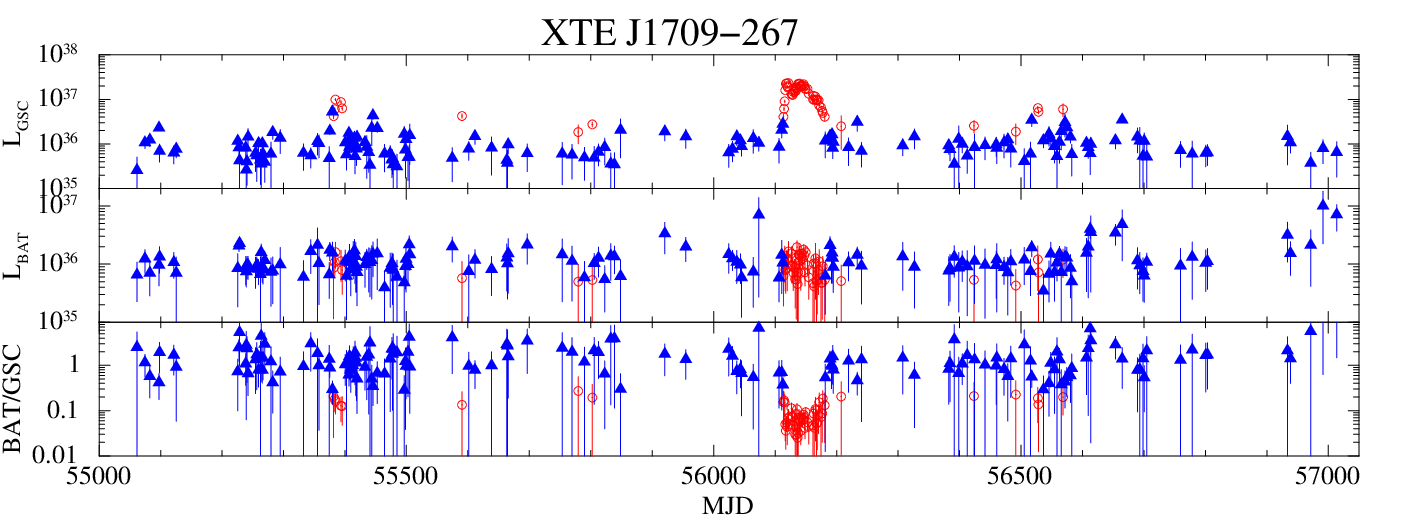}  
\includegraphics[width=16cm]{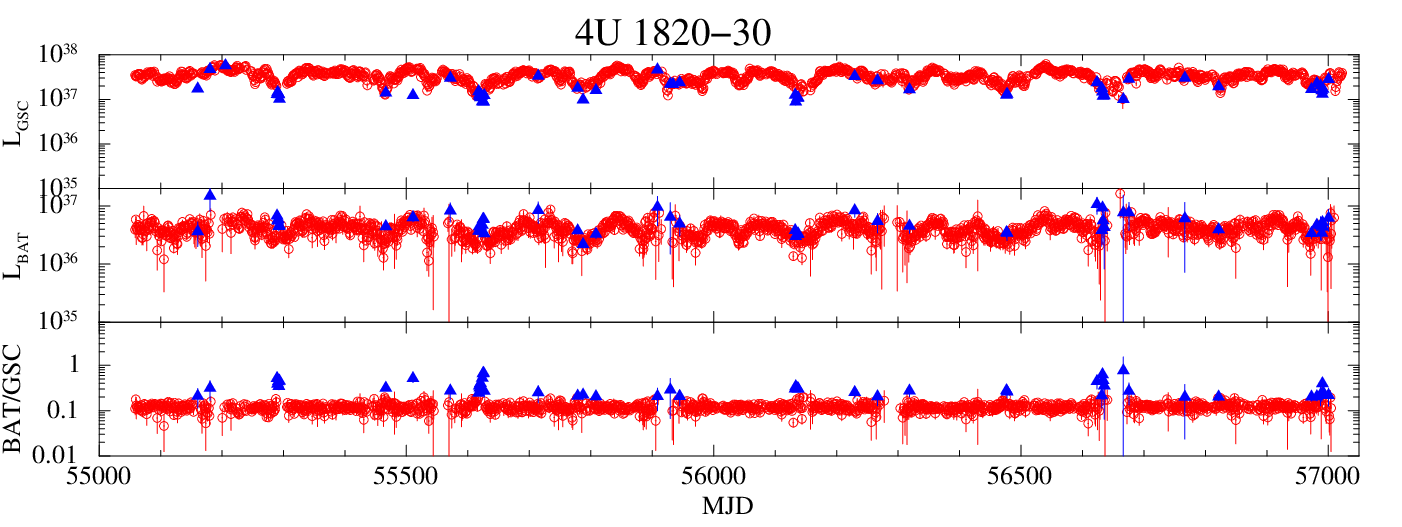}  
 \end{center}
\caption{Same as figure~\ref{fig2}, but for Aql~X-1,
XTE~J1709$-$267, and 4U~1820$-$30. (Color online)}
\label{fig3}
\end{figure*}

We identified the spectral state transition from the BAT$/$GSC hardness ratios
for eleven NS-LMXBs
(4U~1635$-$536, 4U~1705$-$44, 4U~1608$-$52, GS~1826$-$238, Aql~X-1,
XTE~J1709$-$267, 4U~1820$-$30, M~15 X-2, Terzan~5 X-2, Terzan~5 X-3,
and SAX~J1748.9$-$2021).
Among them, five sources
(4U~1820$-$30, M~15 X-2, Terzan~5 X-2, Terzan~5 X-3, and
SAX~J1748.9$-$2021)
are located in globular clusters.
4U~1820$-$30 is the only persistently bright source in the globular cluster, 
and contamination from other X-ray sources 
can be ignored.
However, the other four sources are transients,
whose quiescent states are affected by
contamination.
Therefore, it is difficult to estimate the average luminosity
during quiescence.
In M~15, although the outbursts was identified
with M~15 X-2 \citep{Tomida2013},
X-ray emission in quiescence contains that of
at least two X-ray sources,
4U~2127$+$11 and M~15 X-2 \citep{White2001}. 
In Terzan~5, two outbursts with clear spectral state transition
were detected.
The former outburst was identified with 
Terzan~5 X-2 (IGR~J17480$-$2446: \cite{Ferrigno2010})
and the latter was Terzan~5 X-3
(Swift~J174805.3$-$244637: \cite{Wijnands2012}).
However, the quiescent flux was contributed by at least three sources:
EXO~1745$-$248, Terzan~5 X-2, and  Terzan~5 X-3 \citep{Bahramian2014}.
The outburst of NGC~6440 was identified with
SAX~J1748.9$-$2021 \citep{Suzuki2009}.
However, there are at least two sources SAX~J1748.9$-$2021 and NGC~6440 X-2
\citep{Heinke2010}.
Therefore, the four sources in globular clusters
(M~15 X-2, Terzan~5 X-2, Terzan~5 X-3, and SAX~J1748.9$-$2021)
were excluded from further analysis in this paper. 
The properties of the seven investigated NS-LMXBs 
are summarized in table~\ref{tab1}.

\subsection{Threshold of state transition in the BAT$/$GSC hardness ratio}
Figure~\ref{fig1} shows the histograms (number of days) of
the BAT$/$GSC hardness ratios of the seven NS-LMXBs.
Six of the NS-LMXB exhibit two peaks corresponding to the soft and hard states.
The exception is 4U~1820$-$30, which exists almost exclusively
in the soft state.
Consequently, the dwell-time of the hard state is very short
and its hard-state peak is unclear.
We basically adopted the luminosity of the lowest point between the two peaks
as the threshold between the soft and hard states.
When the lowest points were unclear (Aql~X-1),
we adopted the center value of
the right end of the soft-state distribution and the
left end of the hard-state distribution.
The threshold of 4U~1820$-$30 was determined at the right end of
the Gaussian distribution of the soft state,
where excess points first appear.

The thresholds between the soft and hard states
are summarized in table~\ref{tab2}.
Figures 2 and 3 show the GSC light curves, BAT light curves,
and hardness ratios (BAT$/$GSC) of the seven sources.
In these figures, 
the thresholds separate the data points
into soft and hard states.

\subsection{Mini-outburst and normal outburst}
\label{mini}

\begin{figure*}
 \begin{center}
\includegraphics[width=8cm]{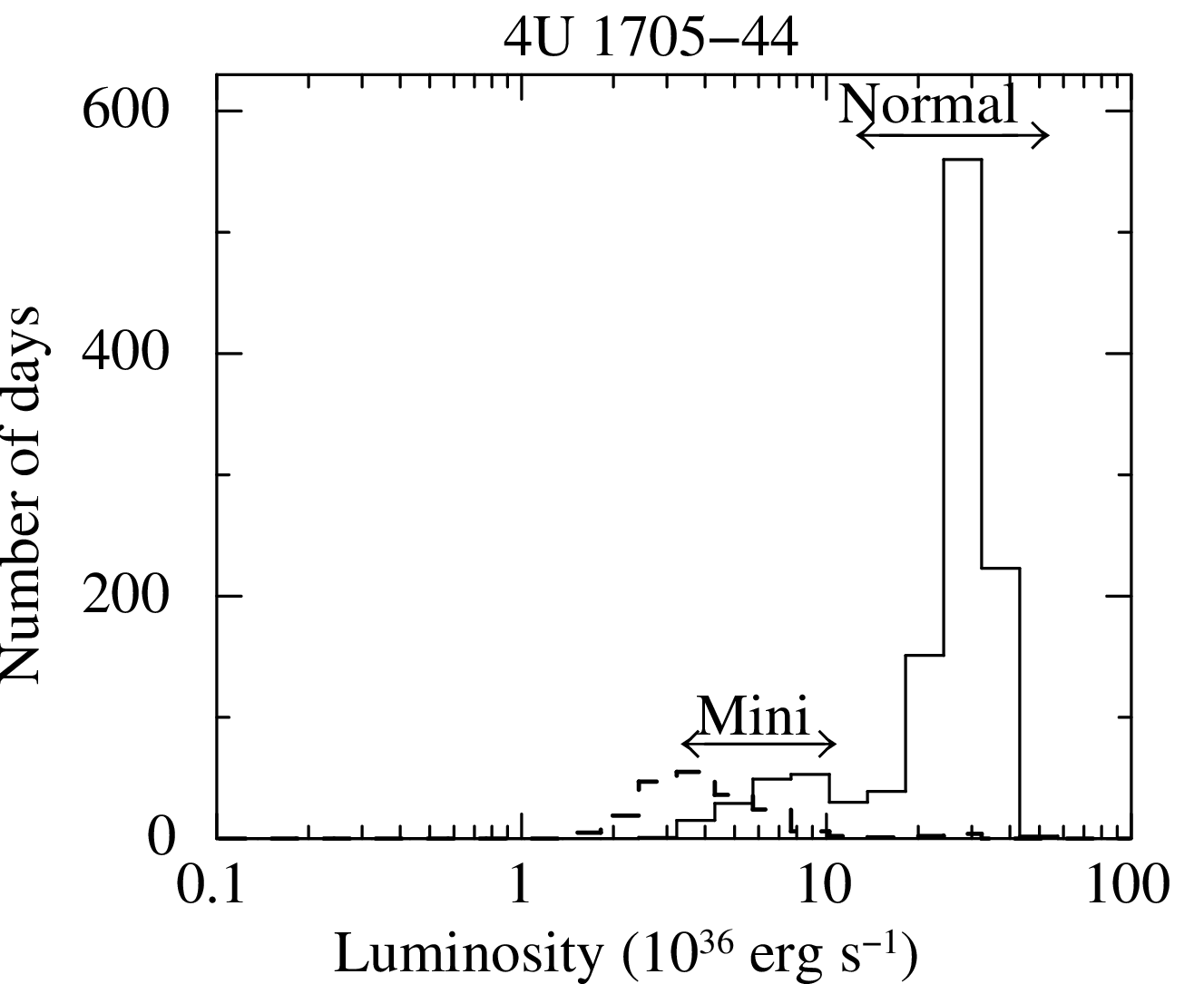} 
\includegraphics[width=8cm]{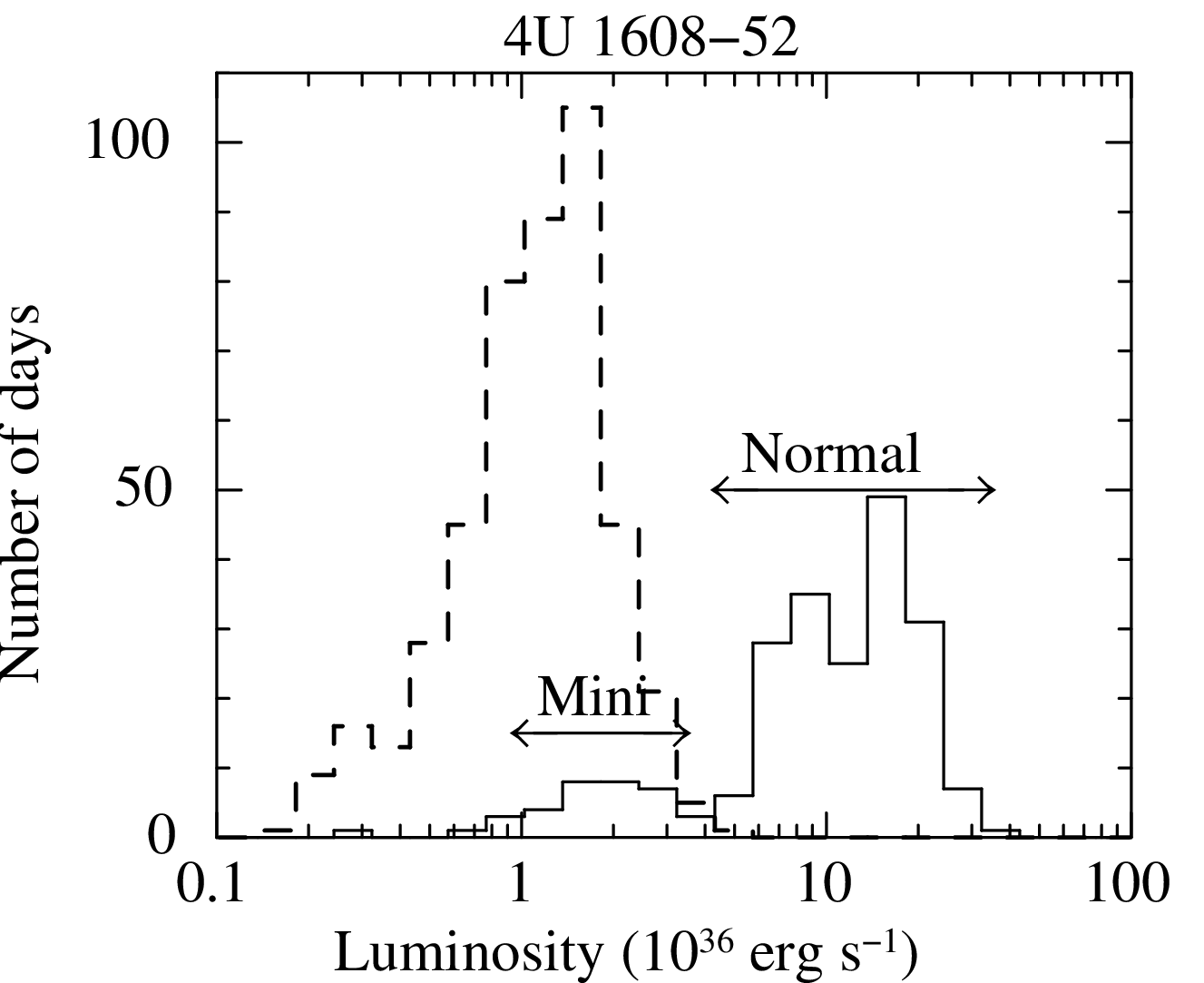} 
 \end{center}
\caption{Luminosity dwell-time distributions of 4U~1705$-$44 and
4U~1608$-$52.
The solid- and dashed-line histograms show the data of soft and hard states,
respectively.
We identified two types of soft states with different luminosities.
Horizontal arrows indicate the ranges of soft states in mini-outbursts and
normal outbursts.
}
\label{fig4}
\end{figure*}

We calculated the average 2--10~keV luminosities of the soft and hard states
using the thresholds of the hardness ratios in table~\ref{tab2}.
Here, the hard state of XTE~J1709$-$267
was obscured by the detection limit of this source
($\sim 10^{36}$ erg s$^{-1}$).
Thus, the calculated average luminosity in the hard state
corresponds to the upper limit.
Two types of soft states with different luminosities
were noticed in 4U~1705$-$44 and 4U~1608$-$52.
The luminosities differ by factors of 3--5 despite
being in the soft state.
In 4U~1705$-$44, the soft-state luminosities were lower
during MJD = 55058--55400 and 55800--56100 than during
other soft-state parts.
Therefore, we calculated the average luminosity of each soft state.
4U~1608$-$52 exhibited a low-luminosity soft state
(below $5\times10^{36}$ erg s$^{-1}$) during MJD = 55900--56100. 
Thus, the average luminosity during MJD = 55900--56100 was separately
calculated in table~\ref{tab2}.
The luminosity dwell-time distributions of 4U~1705$-$44 and 4U~1608$-$52
also exhibit two soft states with different luminosities
(see figure~\ref{fig4}).

In table~\ref{tab2}, the soft states of all sources are
separated into two luminosity classes.
If the average luminosity in the soft state
is below $\sim 10^{37}$ erg s$^{-1}$,
the X-ray variability (repeated small increase) 
is called ``mini-outbursts,''
whereas outburst
with average luminosity above $\sim 10^{37}$ erg s$^{-1}$
is called ``normal outburst.''
For 4U~1820$-$30, the average luminosities of both soft and hard states
are above $\sim 10^{37}$ erg s$^{-1}$.
This source remains in the soft state unless 
the luminosity in the 2--10~keV band decreased
below $\sim 3 \times 10^{37}$ erg s$^{-1}$.
The hard states were identified around 
MJD $\sim$ 55294, 55622, 56135, and 56990.
The X-ray variability cannot be described as an outburst.
Therefore, we categorized it as the ``other,'' 
signifying a different phenomenon.

\begin{figure*}
 \begin{center}
\includegraphics[width=8.2cm]{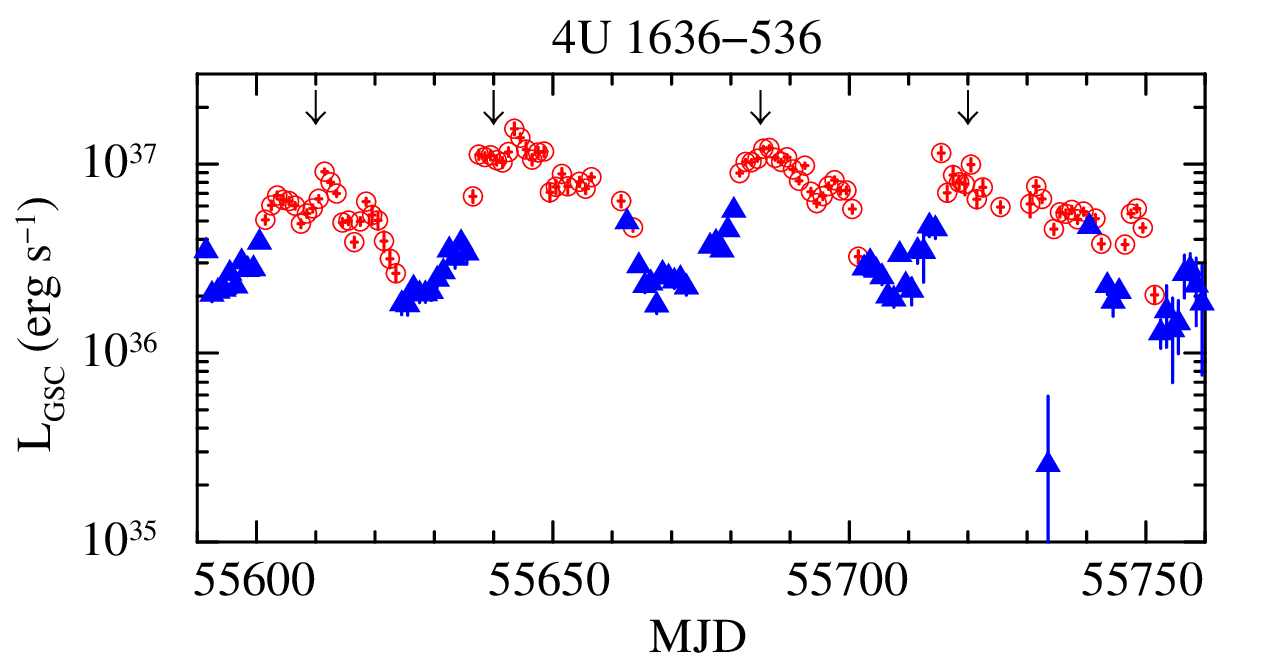} 
\includegraphics[width=8.2cm]{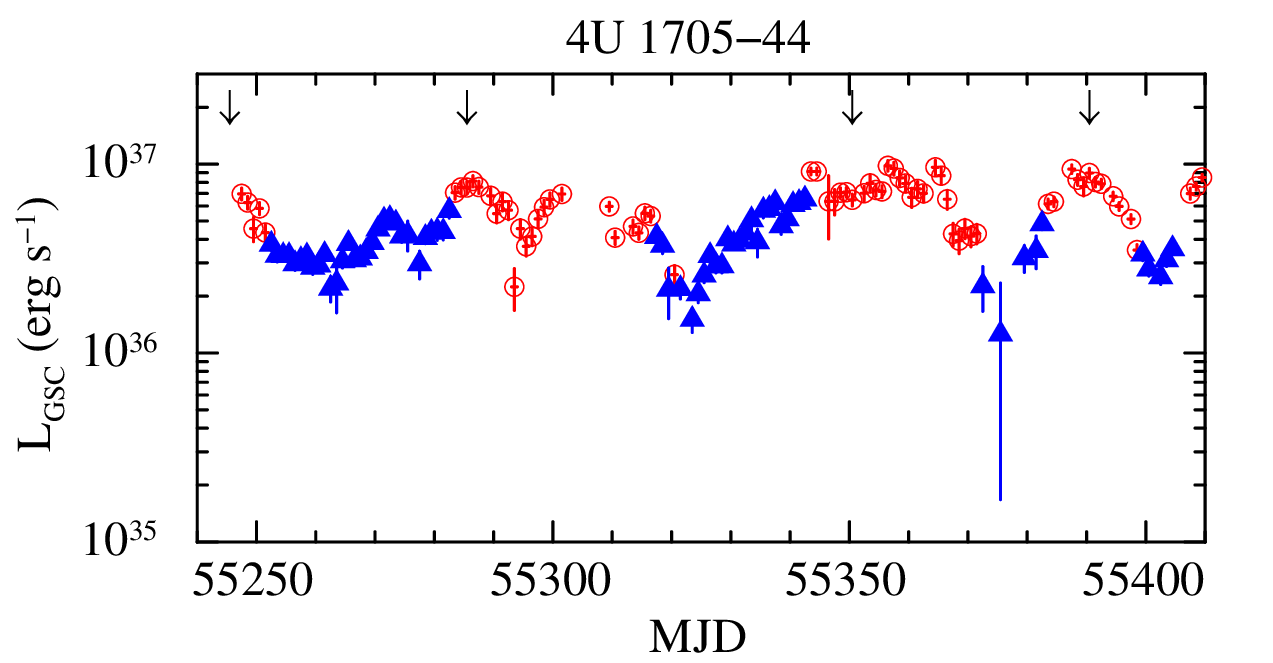} 
\includegraphics[width=8.2cm]{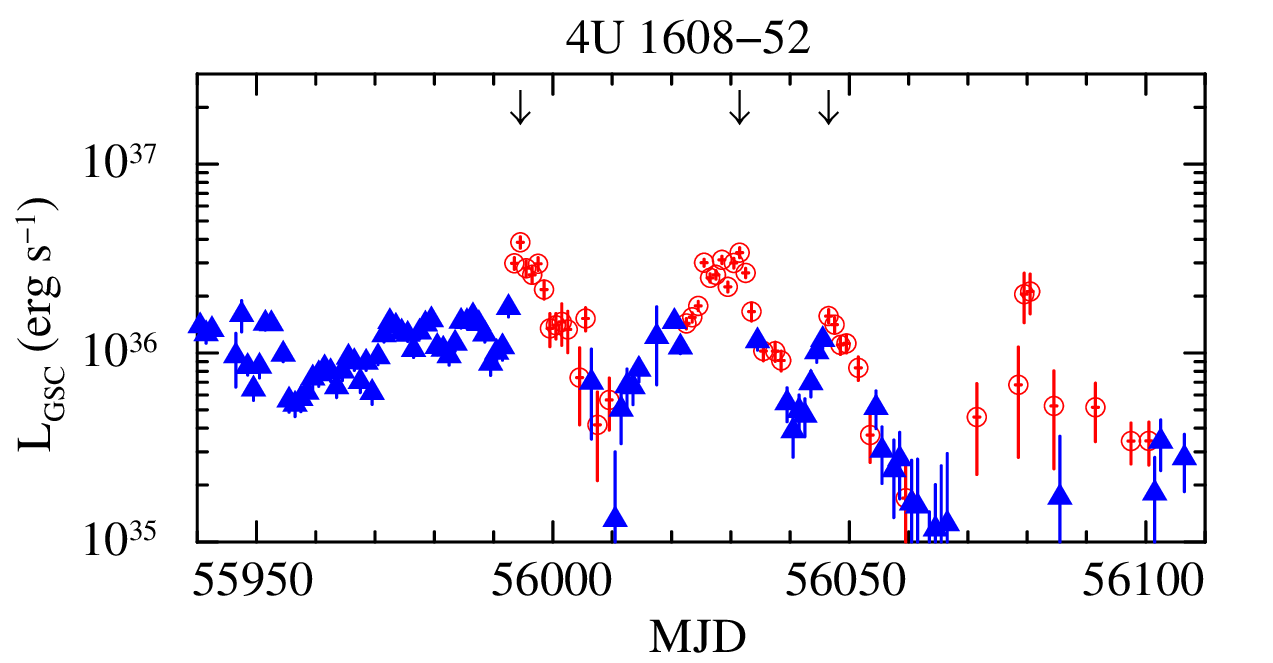} 
\includegraphics[width=8.2cm]{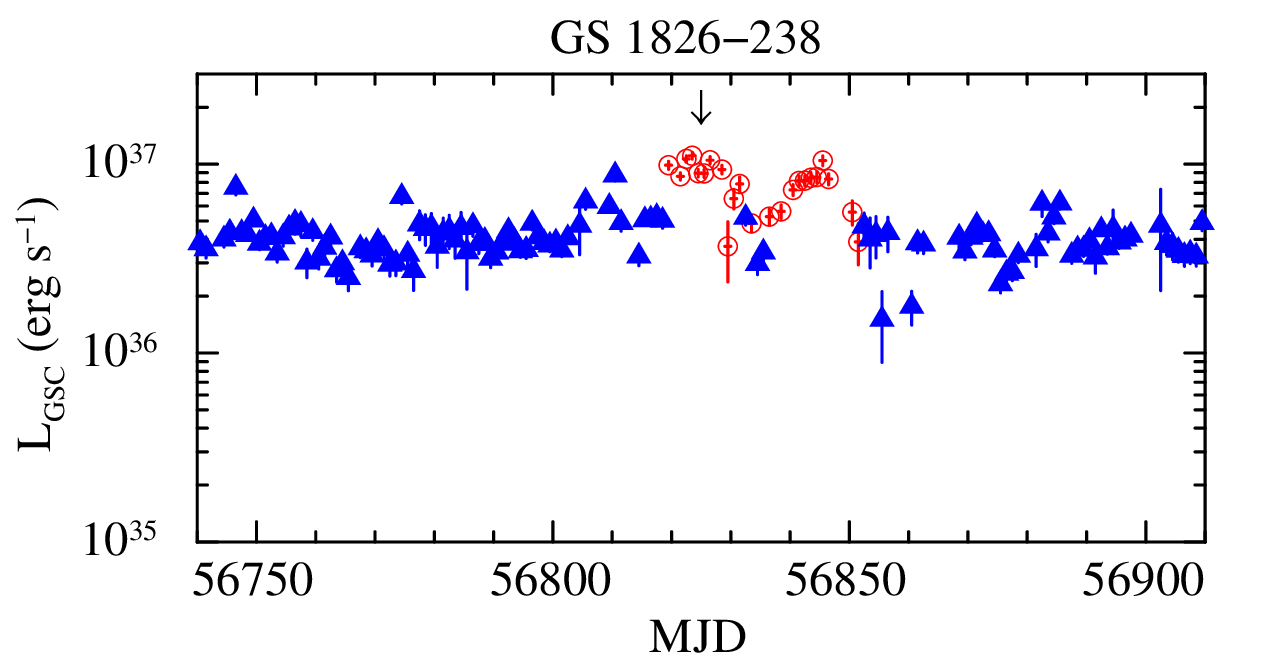} 
 \end{center}
\caption{Magnified light curves of
4U~1636--536, 4U~1705--44, 4U~1608--52, and GS~1826--238.
Mini-outbursts are indicated by arrows.
Red circles and blue triangles represent data of soft and hard states,
respectively. (Color online)}   
\label{fig5}
\end{figure*}

Figure~\ref{fig5} shows typical mini-outburst
from 4U~1636--536, 4U~1705--44, 4U~1608--52, and GS~1826--238.
These plot are magnified regions of their corresponding light curves
in figure~\ref{fig2}.
The individual properties of mini-outburst in the four sources
are summarized as follows:
\begin{description}
\item[4U~1636$-$536:]
Mini-outburst occurred repeatedly
with a period of $\sim$ 30--50~d
while the X-ray luminosity gradually declined
(from $\sim 6 \times 10^{36}$ to $\sim 3 \times 10^{36}$ erg s$^{-1}$
in the 2--10~keV band).
Similar behaviour has been observed since 2002
(Shih et al. 2005, 2011; \cite{Belloni2007}; \cite{Lyu2014}).

\item[4U~1705$-$44:]
Mini-outburst occurred repeatedly
during the periods of no normal outbursts.
The mini-outbursts had shorter durations
than normal outbursts (see figure~\ref{fig2}).
The spectral state transition during the periods
of no normal outbursts
has been also reported by \citet{Homan2009} and
Lin, Remillard, and Homan (2010). 

\item[4U~1608$-$52:]
Four mini-outbursts were identified
on MJD = 55994, 56031, 56046, and 56749
(dates of peak luminosity; see also figure~\ref{fig5}).
The four mini-outbursts occurred at the last period of
long hard state after normal outburst,
where the hard-high state is determined by Matsuoka and Asai (2013).

\item[GS~1826$-$238:]
Mini-outburst occurred
while the X-ray luminosity gradually increased
(from $\sim 2 \times 10^{36}$ to $\sim 5 \times 10^{36}$ erg s$^{-1}$
in the 2--10~keV band).
Although the luminosity in 2--10~keV band
increased by only a few times in the mini-outburst,
the luminosity in the 15--50~keV band decreased
by about one order of magnitude.
Therefore, the spectral state transition was clear.
The spectral state transition has been already
reported by \citet{Nakahira2014}.
\end{description}

\subsection{Determination of transition luminosity}
\label{tranlumi}
\begin{figure*}
 \begin{center}
\includegraphics[width=6cm]{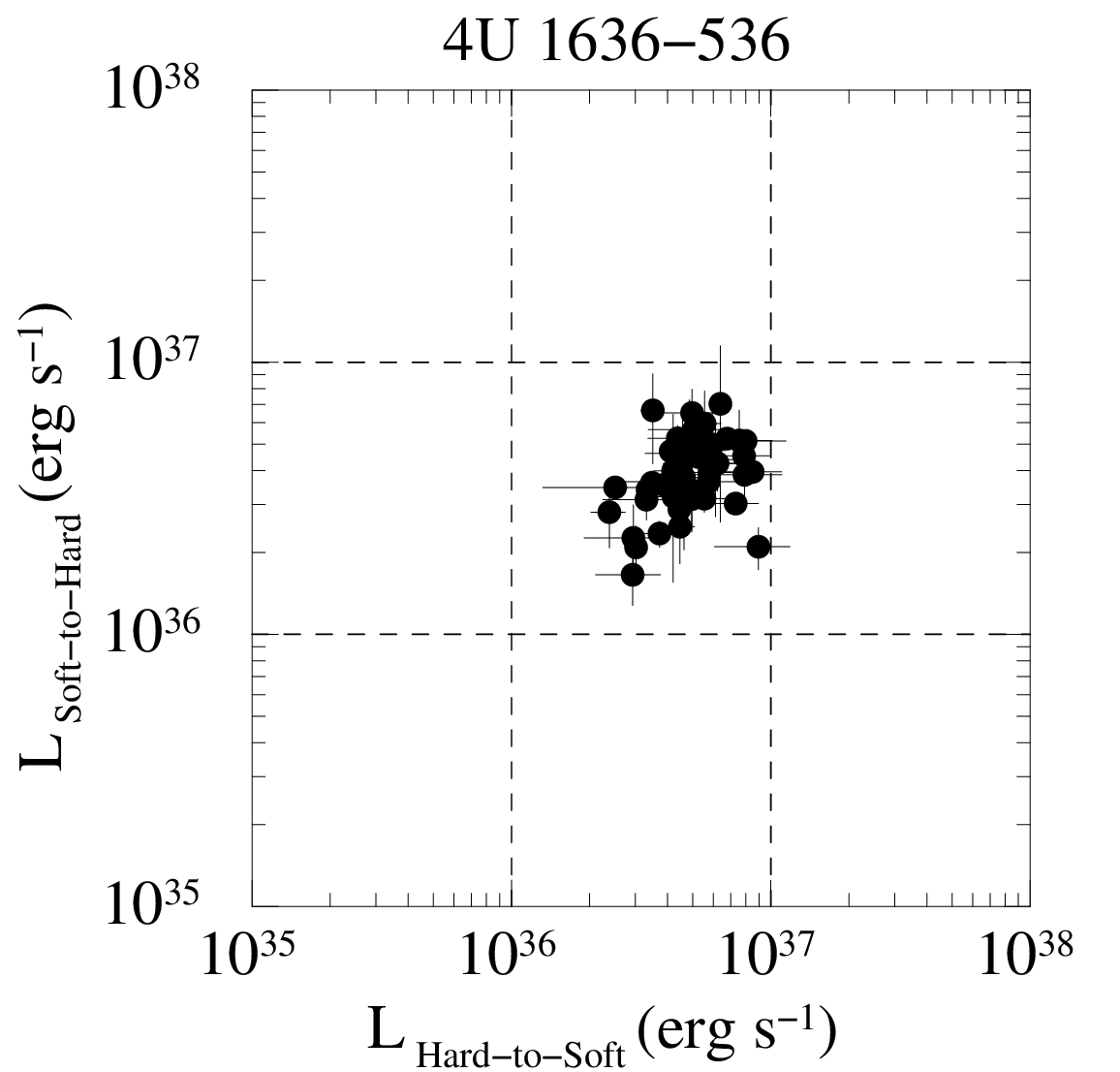}
\includegraphics[width=6cm]{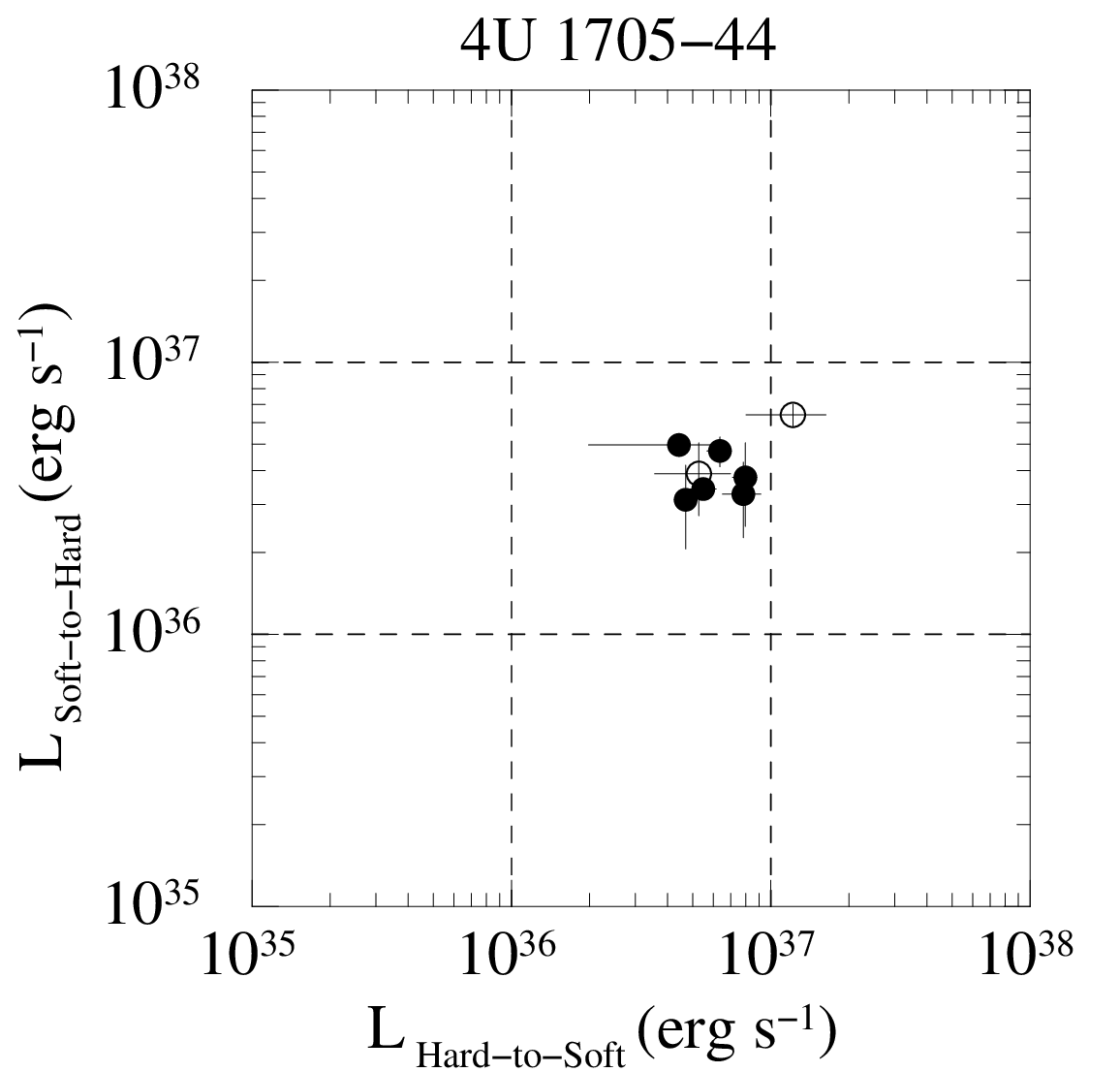}
\includegraphics[width=6cm]{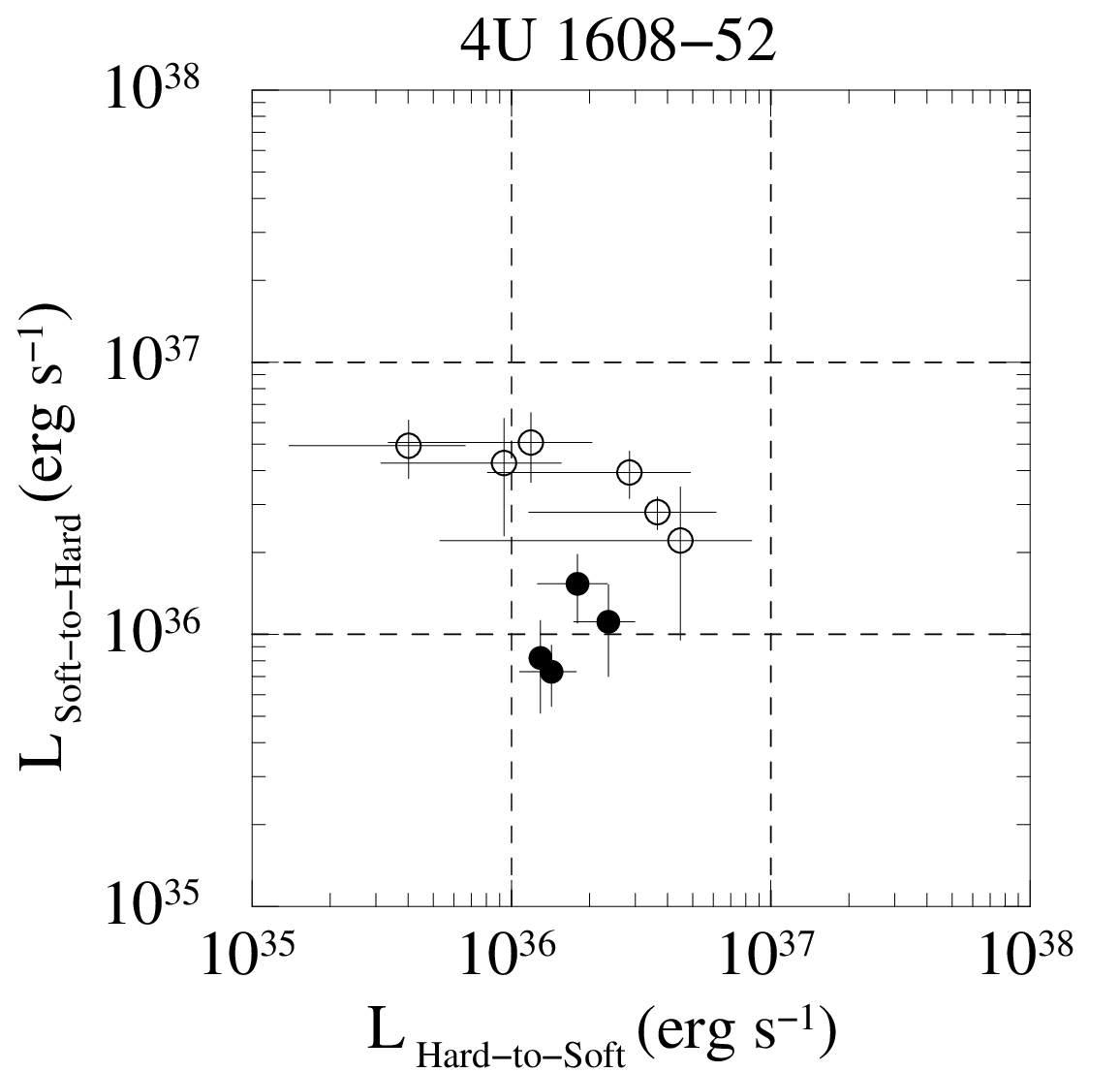}
\includegraphics[width=6cm]{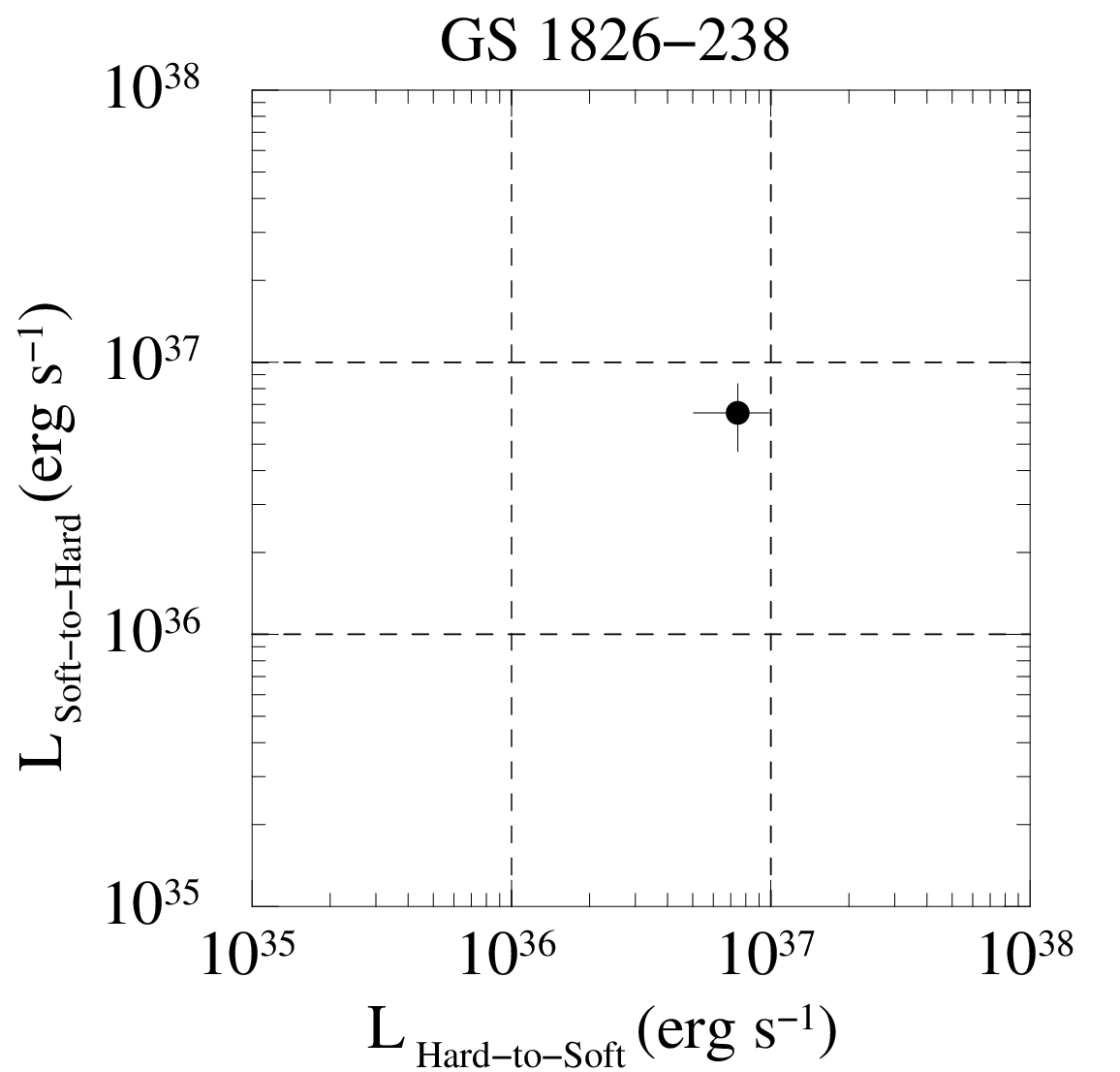}
\includegraphics[width=6cm]{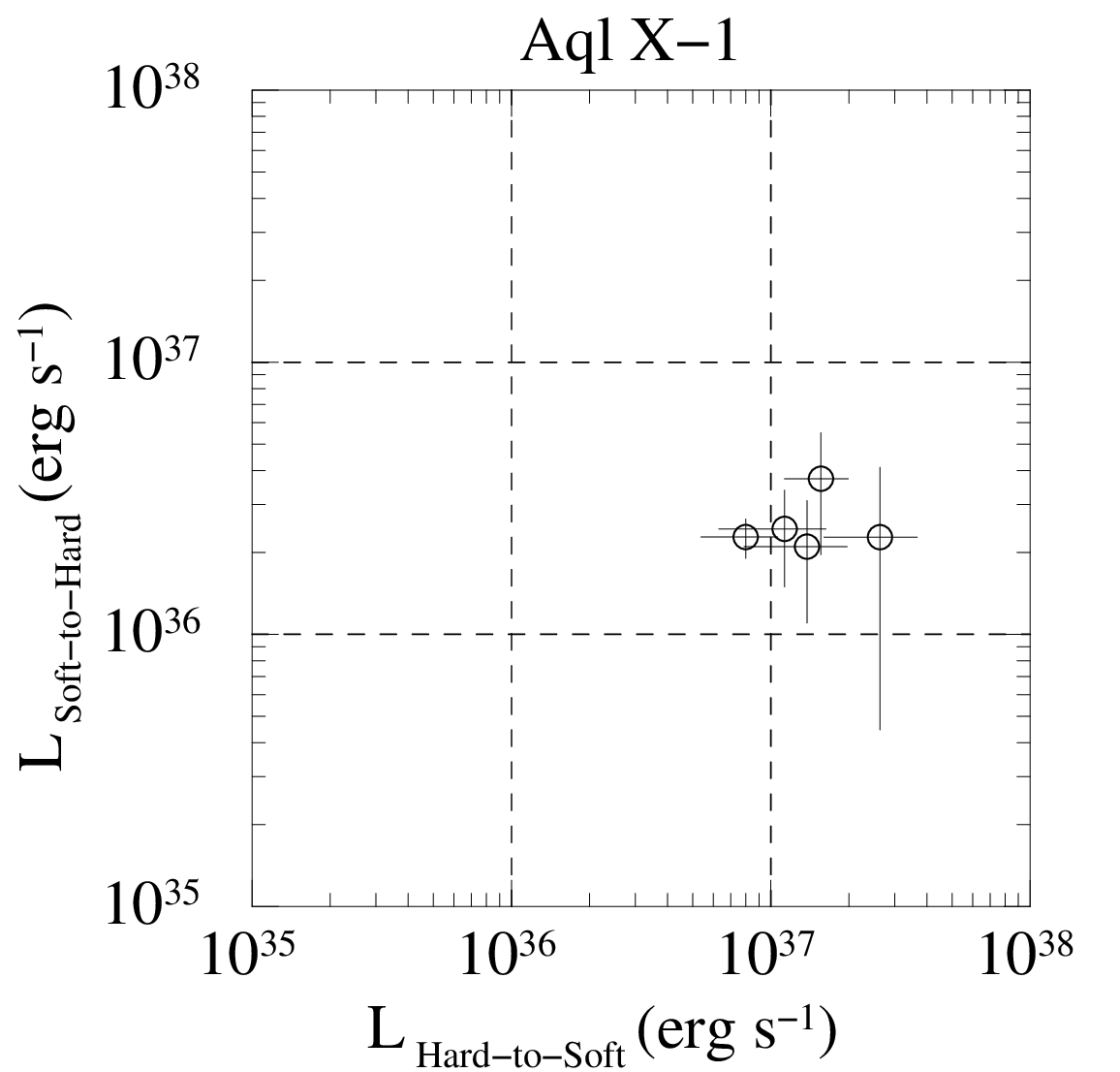}
\includegraphics[width=6cm]{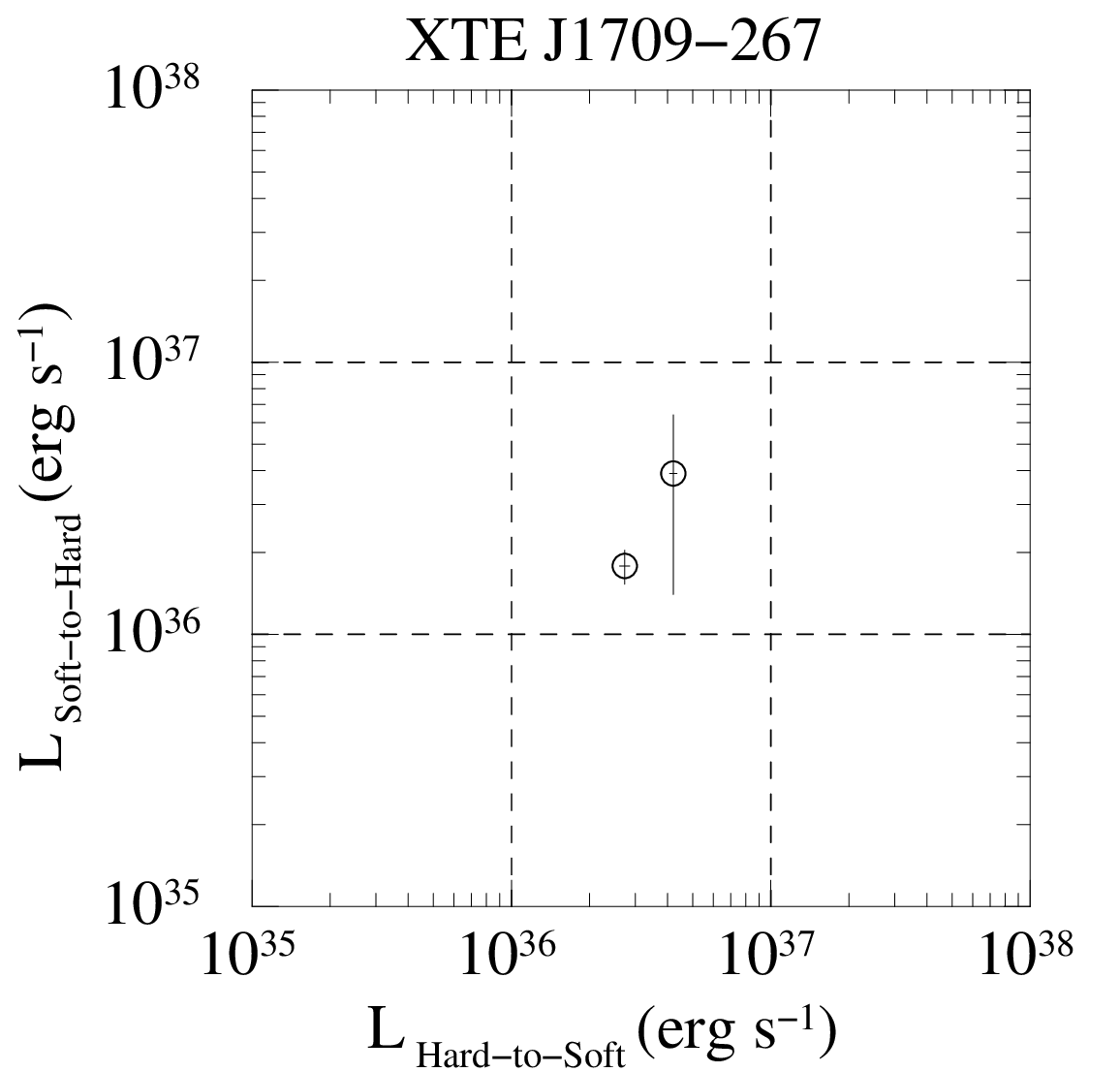}
\includegraphics[width=6cm]{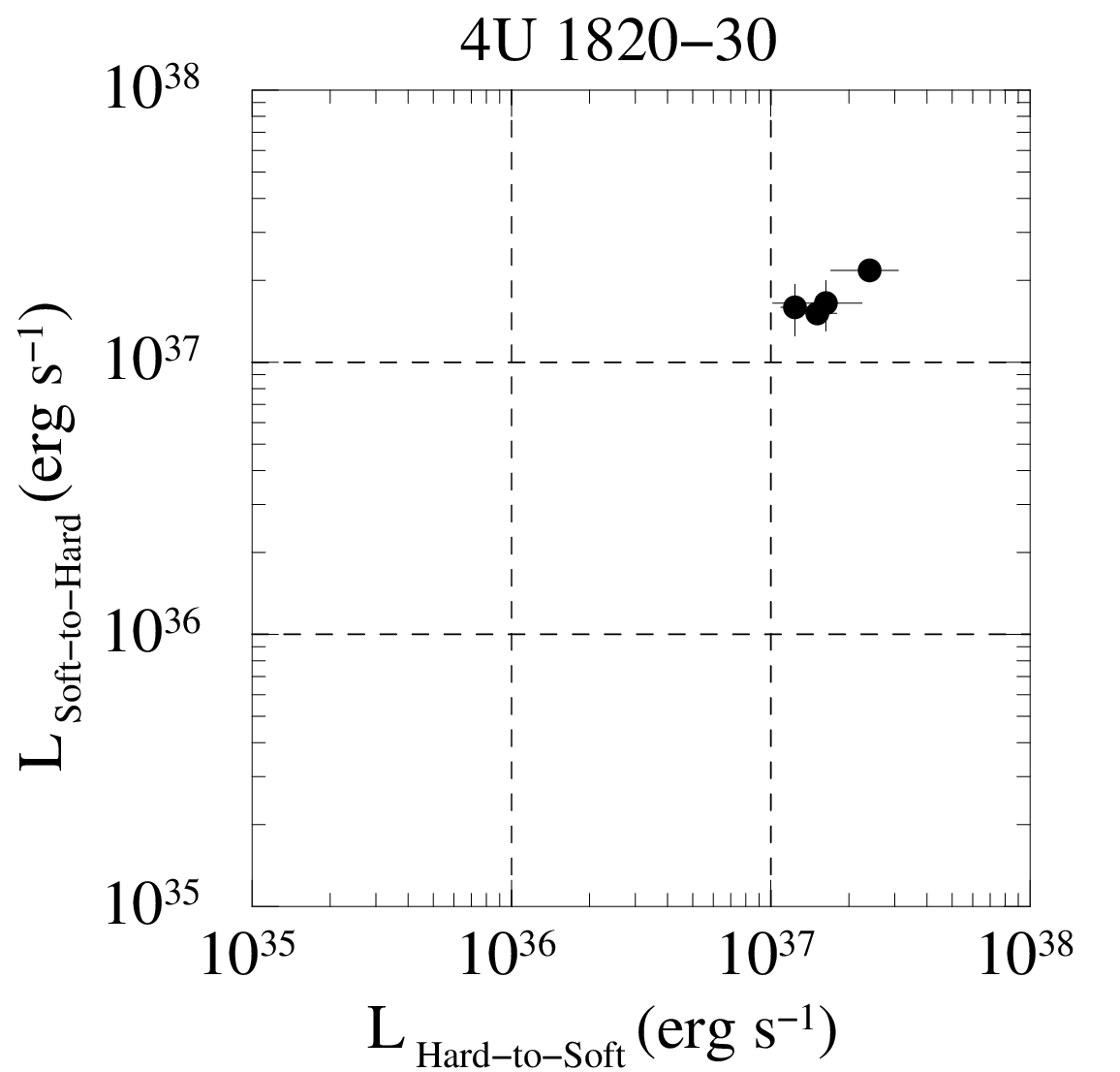}
 \end{center}
\caption{
Transition luminosity (2-10~keV) of hard-to-soft and
soft-to-hard transitions.
Transitions in normal outburst are marked by open circles,
and others (``mini'' and ``other'' in table\ref{tab2}) are marked
by filled circles.}
\label{fig6}
\end{figure*}

We examine the luminosities of hard-to-soft and soft-to-hard
transitions.
Here, we adopt the luminosity in the 2--10~keV band,
because the radiation in this band is dominant in bright NS-LMXBs.
First, we divided the data bins into two states (soft or hard),
using the thresholds of hardness ratio in table~\ref{tab2}. 
A hard-to-soft transition is defined as a soft state continues
for over two days after a hard state and vice versa,
using the data with a significance of more than 1~$\sigma$.
The time and luminosity of the hard-to-soft transition
is estimated by the ``last hard-state'' bin and the ``first soft-state'' bin
using the method described in \citet{Asai2012}.
We employ the center of the two data bins as 
``transition time (the time when transition occurs)'' 
and their time interval as the error on it.
We also define the average luminosity in this time interval
as the ``transition luminosity.'' 
The soft-to-hard transition is similarly defined.
Figure~\ref{fig6} plots the hard-to-soft and
soft-to-hard transition luminosities of the seven LMXBs.
In this figure, the data points of ``mini'' and ``other'' 
(as defined in table~\ref{tab2}) are plotted by ``filled circles,'' 
and are distinguished from the data of normal outburst (open circles).
All transition luminosities of 4U~1820$-$30 categorized as ``other''
are higher than $10^{37}$~erg s$^{-1}$, which is clearly
higher than 4\% of $L_{\rm E}$
even after spectral and bolometric correction as described
in subsection~2.1.
We discuss such high transition luminosities in
subsection~\ref{dis-tranlumi}.

\begin{figure*}
 \begin{center}
\includegraphics[width=8cm]{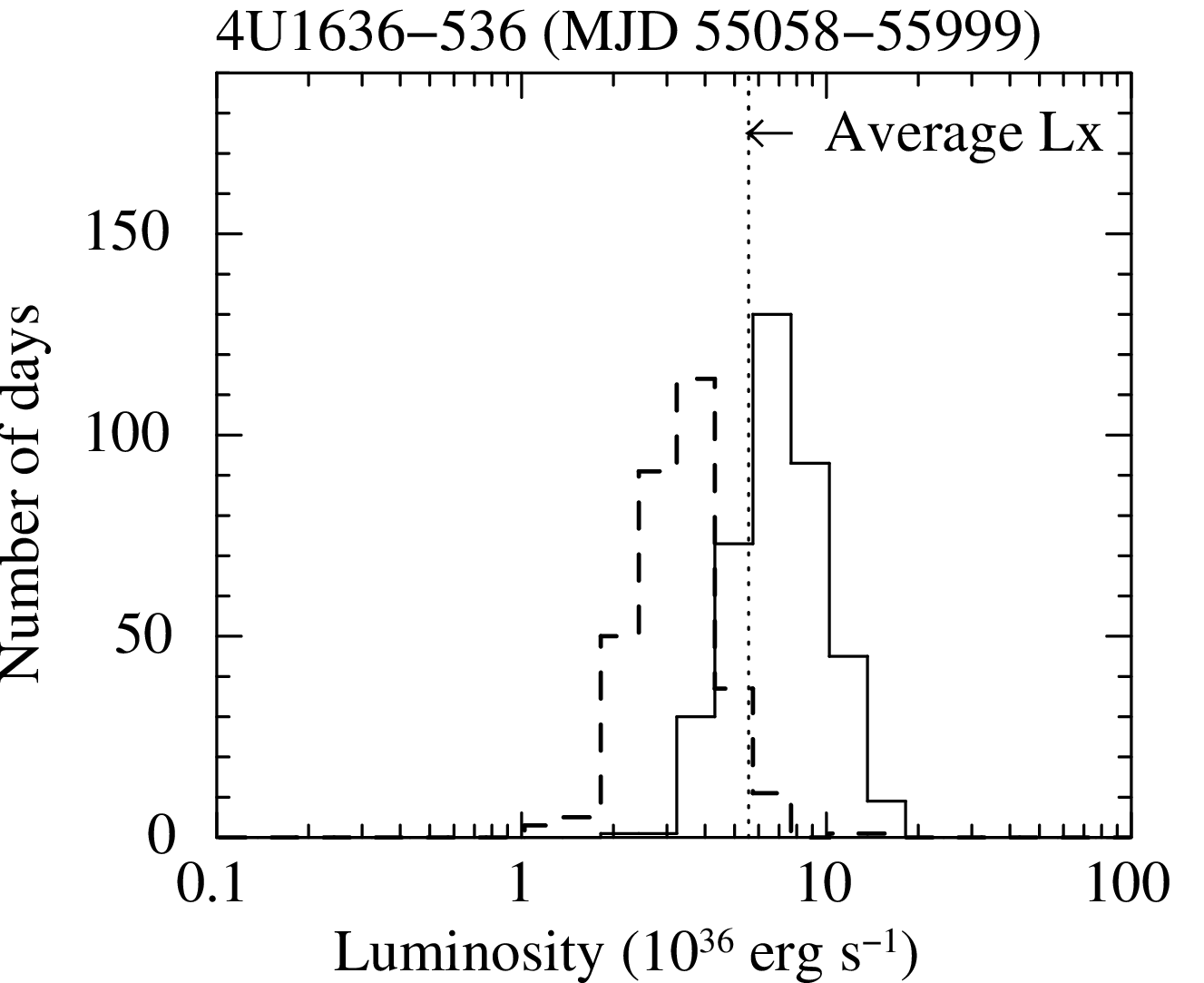} 
\includegraphics[width=8cm]{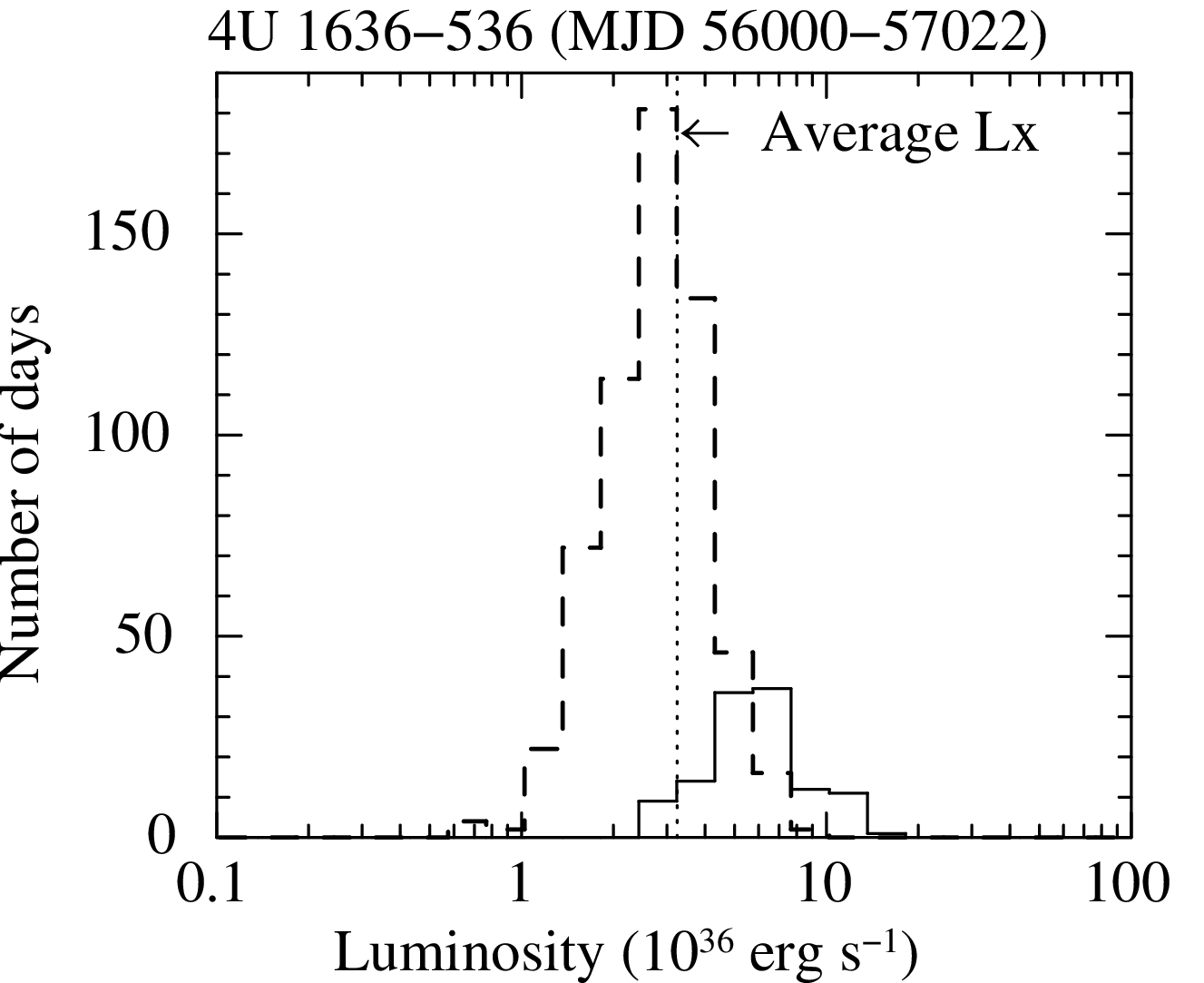} 
 \end{center}
\caption{Luminosity dwell-time distributions of 4U~1636$-$536.
The distributions were constructed from the data 
with a significance of more than 3~$\sigma$.
Left-hand panel displays the distributions of the data for MJD 55058--55999.
Right-hand panel is the data for MJD 56000--57022.
Solid- and dashed line histograms were constructed from data of soft
and hard states, respectively.
The vertical dotted lines indicate the average luminosity 
during each period.}
\label{fig7}
\end{figure*}

Most of transition luminosities of 4U~1636$-$536  
for both hard-to-soft and soft-to-hard transitions are
in the range of (2--10)$\times 10^{36}$~erg s$^{-1}$.
The transition luminosity would be independent of the
luminosity before the transition.
This behaviour was also confirmed in figure~\ref{fig7}.
Here, we divided the observation period into two periods
(MJD = 55058--56000 and MJD = 56000--56899)
and plotted the luminosity dwell-time distribution
during each period.
The average luminosities of the earlier and later periods
were $(5.56\pm0.01)\times10^{36}$~erg s$^{-1}$
and $(3.3\pm0.02)\times10^{36}$~erg s$^{-1}$, respectively.
As the luminosity declined,
the dwell-time of the soft state became shorter, whereas
that of the hard state became longer.
This means that the transition luminosity remains almost constant.

In five normal outbursts 
(2011-Jun of 4U~1705$-$44, 2010-Sep, 2011-Oct, 2013-Jun, and
2014-Jul of Aql~X-1), 
the hard-to-soft transition luminosities are 
higher than $10^{37}$~erg s$^{-1}$, which is
higher than 4\% of $L_{\rm E}$.
These outbursts with higher hard-to-soft transition luminosities
are classified as S-type defined by \citet{Asai2012}.

In a normal outburst of 4U1608$-$52 (2010-Mar), 
the hard-to-soft transition luminosities are lower than
$10^{36}$~erg s$^{-1}$, which would be
1\% of $L_{\rm E}$.
However,
the data bins of the first soft state 
in the one-day BAT light curve
have relatively large uncertainties. 
For example, at a significance level exceeding 2~$\sigma$,
no data bins can be assigned to the first soft state,
because of large data gaps.
The lowness of the transition luminosity
depends on the data quality, so is not discussed here.
The time and luminosity of the state transitions and 
the soft-state durations of 
normal outbursts from 4U~1705$-$44, 4U~1608$-$52,
Aql~X-1, and XTE~J1709$-$267 are summarized
in table~\ref{tab3}.

\begin{table*}
\caption{Summary of normal outbursts from 4U~1705$-$44, 4U~1608$-$52, Aql~X-1,
and XTE~J1709$-$267}
\label{tab3}
\begin{center}
\begin{tabular}{lccccc}
\hline
Normal & \multicolumn{2}{c}{Hard-to-soft transition}
& \multicolumn{2}{c}{Soft-to-hard transition} & Duration \\
outburst & Time (MJD) & Luminosity\footnotemark[$*$]
&  Time (MJD) & Luminosity\footnotemark[$*$] & 
of soft state (d)\\
\hline
\multicolumn{6}{c}{4U~1705$-$44} \\
\hline
2010-Jul & $55406.0\pm1.5$ & $5.3\pm1.7$ & $55695.5\pm2.0$ & $4.0\pm1.1$ & $289.5\pm3.5$ \\
2011-Jun & $55716.0\pm0.5$ & $12.2\pm4.1$ & $55921.5\pm1.0$ & $6.4\pm0.6$ & $205.5\pm1.5$ \\
2012-Apr & $56039.0\pm0.5$ & $6.6\pm0.3$ & --\footnotemark[$\dagger$]& --\footnotemark[$\dagger$] & --\footnotemark[$\dagger$] \\
\hline
\multicolumn{6}{c}{4U~1608$-$52} \\
\hline
2010-Mar & $55254.5\pm3.0$ & $0.4\pm0.3$ & $55273.0\pm0.5$ & $4.9\pm1.2$ & $18.5\pm3.5$ \\
2011-Mar & $55644.0\pm0.5$ & $3.7\pm2.5$ & $55729.5\pm2.0$ & $2.8\pm0.4$ & $85.5\pm2.5$ \\
2012-Oct & $56209.0\pm1.5$ & $0.9\pm0.6$ & $56228.5\pm1.0$ & $4.3\pm2.0$ & $19.5\pm2.5$ \\
2013-May & $56441.0\pm0.5$ & $2.8\pm2.0$ & $56457.0\pm0.5$ & $3.9\pm0.8$ & $16.0\pm1.0$ \\
2014-Jan & $56672.0\pm1.5$ & $4.5\pm4.0$ & $56741.5\pm3.0$ & $2.2\pm1.3$ & $69.5\pm4.5$ \\
2014-Oct & $56935.0\pm0.5$ & $1.2\pm9.0$ & $56945.0\pm0.5$ & $5.1\pm1.1$ & $10.0\pm1.0$ \\
\hline
\multicolumn{6}{c}{Aql~X-1} \\
\hline
2009-Nov & $55153.0\pm0.5$ & $8.0\pm3.0$ & $55171.0\pm0.5$ & $2.3\pm0.4$ & $18.0\pm1.0$ \\
2010-Sep & $55451.5\pm1.0$ & $15.6\pm4.3$ & $55478.0\pm1.5$ & $3.7\pm1.8$ & $26.5\pm2.5$ \\
2011-Oct & $55856.0\pm0.5$ & $13.8\pm5.9$ & $55906.0\pm1.5$ & $2.1\pm1.0$ & $50.0\pm2.0$ \\
2013-Jun & $56459.0\pm0.5$ & $26.3\pm10.3$ & $56513.0\pm7.5$ & $2.3\pm1.8$ & $54.0\pm8.0$ \\
2014-Jul & $56582.0\pm0.5$ & $11.3\pm5.0$ & $56874.0\pm0.5$ & $2.4\pm0.9$ & $22.0\pm1.0$ \\
\hline
\multicolumn{6}{c}{XTE~J1709$-$267} \\
\hline
2010-Jul & $55380.5\pm1.0$ & $4.8\pm0.6$ & $55398.0\pm2.5$ & $3.7\pm2.6$ & $17.5\pm3.5$ \\
2012-Aug & $56113.0\pm0.5$ & $3.4\pm0.6$ & $56181.0\pm0.5$ & $2.6\pm1.5$ & $68\pm1$ \\
\hline
\multicolumn{6}{@{}l@{}}{\hbox to 0pt{\parbox{140mm}{\footnotesize
\par\noindent
\footnotemark[$*$]
Luminosity in the 2--10~keV band in the unit of $10^{36}$~erg~s$^{-1}$.
\par\noindent
\footnotemark[$\dagger$]
The outburst was continuing at the end of this observation (2015.12.31)
\par\noindent
}\hss}}
\end{tabular}
\end{center}
\end{table*}

\section{Discussion}

\subsection{Transition luminosity}
\label{dis-tranlumi}

We investigated the luminosities in the 2--10~keV band
of hard-to-soft
and soft-to-hard transitions in seven sources.
There are two cases in which 
the transition luminosities are higher than $10^{37}$~erg s$^{-1}$,
which is higher than 4\% of $L_{\rm E}$.
The one is the luminosities of S-type hard-to-soft transitions
\citep{Asai2012}, 
which are higher than 4\% of $L_{\rm E}$.
According to \citet{Asai2012}, 
X-ray irradiation before the outbursts
might heat up the accretion disk and delay the optically thin-to-thick
disk transition.
In this scenario, the optically thin-to-thick disk transition
would be suppressed by the large Compton cloud formed
by the X-ray irradiation.
The S-type outbursts identified in this study
were five normal outbursts from 4U~1705$-$44 and Aql~X-1.

The other
is the transition luminosities of 4U~1820$-$30.
Such high transition luminosities are similar to the hard-to-soft
transitions of S-type in normal outburst.
Both soft and hard states exhibited high average luminosity
in the 15--50~keV band
[soft state: $(4.07\pm0.02)\times10^{36}$ erg s$^{-1}$,
hard state: $(5.3\pm0.3)\times10^{36}$ erg s$^{-1}$].
It is suggested that the Compton cloud is large and
an optically thin disk could exist at such high luminosity. 
\citet{Titarchuk} analyzed the RXTE data of 4U~1820--30
over a similar epoch of luminosity decrease to our data analysis.
They reported that
a dramatic rise in the electron temperature of the Compton cloud
(from 2.9 to 21~keV) and a state transition
from upper banana to island.
The results support that the optically thick-to-thin disk transition
occurs at such high luminosity.

In summary, the spectral state transition usually corresponds to the state
transition of the inner accretion disk. 
However, spectral state transition might result from
changes in not only the disk state but also the 
electron temperature of the Compton cloud.

\subsection{Properties of mini-outburst}

\begin{figure}
 \begin{center}
  \includegraphics[width=7cm]{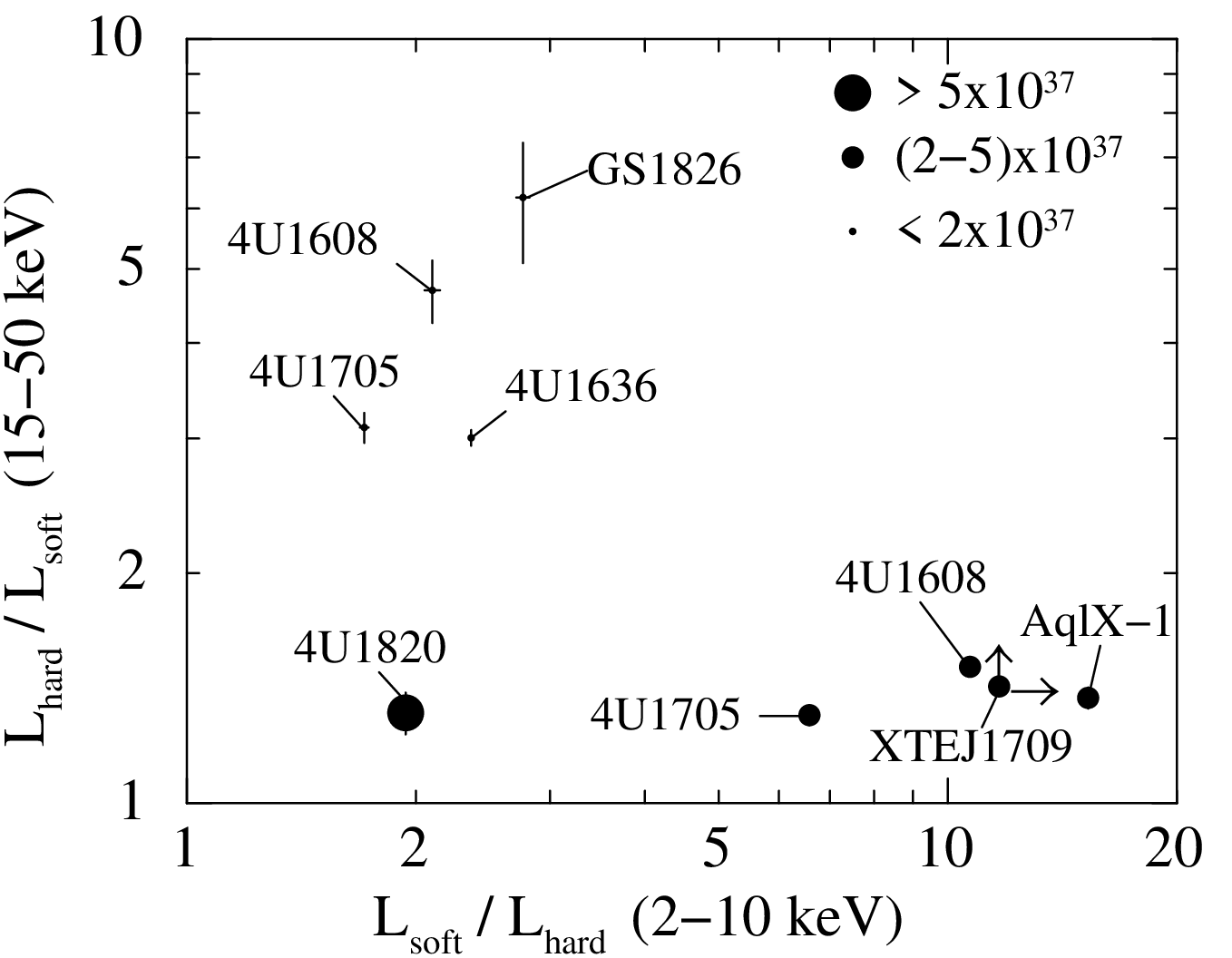} 
 \end{center}
\caption{
Average luminosity ratios of the soft and hard states
in the 2--10~keV band (GSC)
and the 15--50~keV band (BAT).
Mark size represents the peak luminosity of the soft state.
The data point of XTE~J1709$-$26 is a lower limit
because of the detection limit during the period of no outbursts.}
\label{fig8}
\end{figure}

\begin{figure}
 \begin{center}
  \includegraphics[width=8cm]{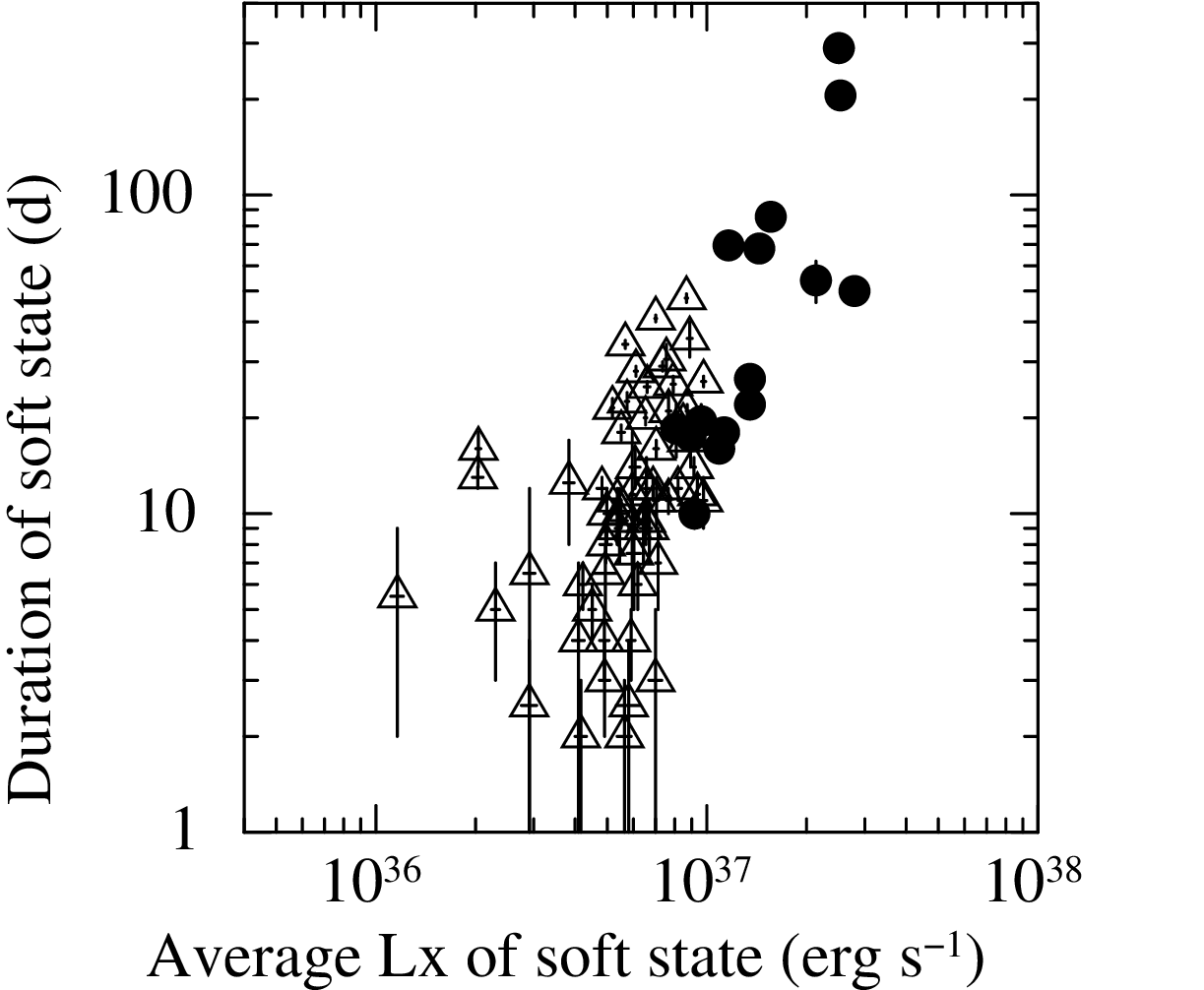}
 \end{center}
\caption{Duration vs.~average luminosity of the soft states
in the 2--10~keV band.
Filled circles and open triangles represent data
of normal outbursts and mini-outbursts,
respectively.
The data of 4U~1820--30 were excluded from this figure,
because the behaviour is different from that of both normal outburst
and mini-outburst
as described in table~\ref{tab2} and subsection~\ref{mini}.}
\label{fig9}
\end{figure}

As described in subsection~\ref{mini}, 
the X-ray variability with spectral state transition
was divided into mini-outburst and normal outburst by
the average luminosity of the soft state in the 2--10~keV band.
Moreover, in the 15--50~keV band, the ratio of the average luminosities
in the soft and hard states also differed between
mini-outbursts and normal outbursts.
Figure~\ref{fig8} shows the ratios of the average luminosities
of the soft and hard states
in the 2--10~keV band and that in the 15--50~keV band.
The ratio in the 15--50~keV band is larger in the mini-outbursts
than in normal outbursts.
The difference is relevant to the duration of the soft state.
Figure~\ref{fig9} plots the duration versus the average luminosity
of the soft states.
Here, we excluded the data of 4U~1820--30,
which behave differently from both normal outburst
and mini-outburst as described
in table~\ref{tab2} and subsection~\ref{mini}.
The mini-outbursts tend to have shorter durations
than normal outbursts.
In both mini-outburst and normal outburst, 
the luminosity in the 15--50~keV band drops once
at the hard-to-soft transition,
then increases again.
The post-transition increase is caused by the increase of seed photons 
(see figure~8 in \cite{Asai2012}).
However, in the short duration of mini-outburst,
the 15--50~keV luminosity
does not recover to its hard-state level just before the
transition.
Therefore, the luminosity ratio of the soft and hard states
becomes large in the 15--50~keV band. 

The properties of mini-outburst are summarized as follows:
(1) mini-outburst has smaller amplitude and
shorter duration than those of normal outburst,
and (2) most of the transition luminosities of mini-outbursts are
in the range of 1\%--4\% of $L_{\rm E}$.
Therefore, both mini-outburst and normal outburst
would be similarly interpreted by the disk instability model.
The time scale of the mini-outburst supports
the disk instability model.
Following \citet{Frank1992}, we estimated the viscous time scale
in the outer disk. 
The viscous time scale was given by
\begin{equation}
t_{\rm visc}\sim 3 \times 10^5~\alpha^{-4/5}\,
\dot{M}_{16}^{-3/10} M_{\rm NS}^{1/4} \ (R_{\rm d}/10^{10})^{5/4}~\rm s,
\end{equation}
where
$\dot{M}_{16}$ is the mass accretion rate in units of $10^{16}$ g s$^{-1}$,
$M_{\rm NS}$ is the neutron star mass of 1.4$\Mo$, and 
$R_{\rm d}$ is the outer radius of the accretion disk.
Assuming that 
$\alpha$ = 0.1--1,
$\dot{M}_{16}$ = 5.3 ($L_{\rm x} \sim 10^{37}$~erg s$^{-1}$,
the maximum luminosity of a mini-outburst),
and
$R_{\rm d} \sim 3\times 10^{10}$~cm
(the $R_{\rm d}$ of 4U~1636--536; see figure~\ref{fig9}
in the next subsection),
we obtained $t_{\rm visc}\sim$ 10--60~d.
This time scale is roughly comparable with the typical time scale of
a mini-outburst.
Furthermore, 
Mineshige and Osaki (1985) predicted
a small-amplitude variability
(resembling a mini-outburst)
in the disk instability model, 
although their calculation assumed a dwarf nova.
We discuss the relation of the mini-outburst and the disk instability model
in the next subsection.

The periodic variability observed in KS~1731$-$320
is similar to the mini-outbursts in 4U~1636$-$536,
although the presence of spectral state transitions
is unclear.
According to Revnivtsev and Sunyaev (2003),
KS~1731$-$320 exhibited strong variability
on a monthly time scale during MJD $\sim$ 51100--52000.
They interpreted the periodic variability as disk precession.
The luminosity in the 2--10~keV band obtained from the public data
of ASM on RXTE, was between $\sim 3 \times 10^{36}$~erg s$^{-1}$
and $\sim 1 \times 10^{37}$~erg s$^{-1}$, 
assuming a distance of 7~kpc \citep{Muno2000}. 
The luminosity range of the variability is close to
1\%--4\% of $L_{\rm E}$.
Therefore, we suggest that the variability could result from
a repeat of mini-outburst.

\citet{Simon2010} referred the variability in KS~1731$-$320
that follows the main outburst as  ``echo outbursts.''
They suggested that echo outbursts are triggered by 
the configuration of the inner ionized stable region
surrounded by the outer unstable region.
They also pointed out the important role of
thermal instability of the accretion disk
in the luminosity decline.
\citet{Shih2005} also reported that the variability
during the X-ray decline in both 4U~1636$-$536 and KS~1731$-$260
was similar.
They suggested the relation of the variability and
the thermal instability of an accretion disk.

\subsection{Mini-outburst and disk instability model}

Normal outbursts in NS-LMXBs are usually interpreted by the disk instability
model like the outburst of dwarf novae. 
The theoretical disk instability model of 
Mineshige and Osaki (1983, 1985)
predicts both large-amplitude outbursts (roar type) and
small-amplitude variability (purr type).
The former is considered to correspond to normal outburst,
but no correspondence has been found for the latter.
Here, we propose that purr-type outbursts
correspond to mini-outbursts, because the characteristics
(small-amplitude and short duration) of both types are quite similar.

We now discuss the possibility of the purr-type outburst.
Most of the transition luminosities of mini-outbursts
were in the range of 1\%--4\% of $L_{\rm E}$.
It would indicate that the spectral state transitions
are due to the inner disk transitions, which 
are caused by thermal instability of the outer disk.
The main difference of the two-types instability
(roar type and purr type) is
in the assignment of $\alpha$ parameter. 
The $\alpha$ parameter is usually considered
to depend on the disk temperature, 
although there are several theoretical models
(e.g., \cite{Hameury1998}).
In order to explain a normal outburst,
a high $\alpha$ is assigned in a hot state (upper branch in S-shape curve)
and a low $\alpha$ is assigned in a cool state (lower branch).
As a result, the S-shape curve is modified and 
the width of the middle branch is extended.
This means that the instability can propagate through the whole disk
(roar type).
On the other hand, when the same $\alpha$ is assigned in 
both a hot and cool state (upper and lower branch),
the width of the middle branch is narrow, and then
the instability cannot propagate through the whole disk (purr type).

Here, we examine the influence of the irradiation
on the S-shaped curve.
Tuchman, Mineshige, and Wheeler (1990) calculated
the instability under irradiation and
generated the S-shaped curves at different irradiation temperatures
(see figure~1 in \cite{Tuchman1990}).
There are two modifications of the S-shaped curve 
due to the irradiation.
One is to move the lower branch upward in the diagram
and
the other is to shorten the width of the middle branch.
Those two lead to that the $\alpha$-value of the lower branch
becomes close to that of the upper branch, and then
the approximation of a constant-$\alpha$ disk
would be suitable.

\begin{figure}
 \begin{center}
  \includegraphics[width=8cm]{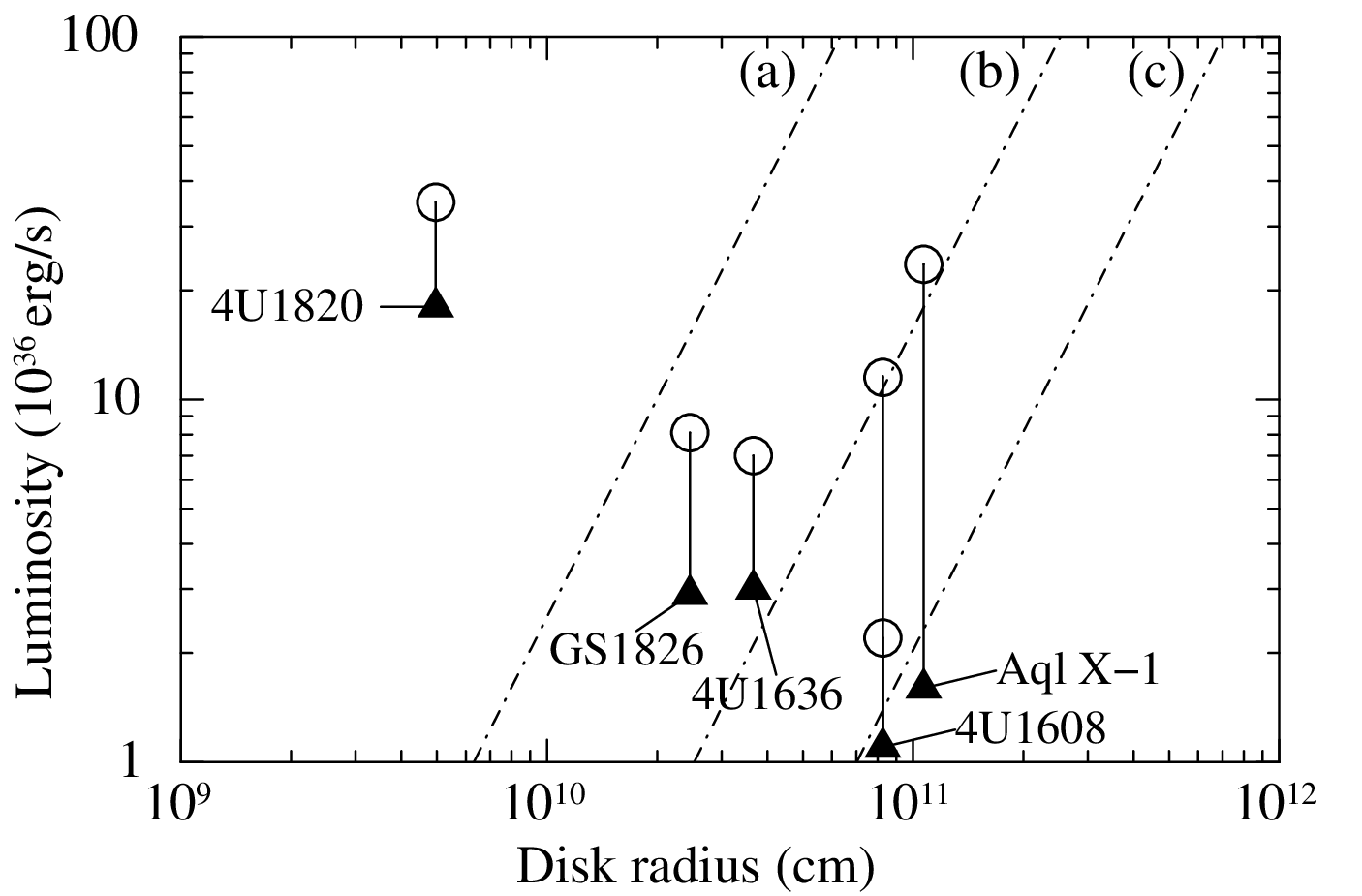} 
 \end{center}
\caption
{Average X-ray luminosity in the 2--10~keV band
as a function of outer disk radius.
Open circles and filled triangles represent data of soft and hard states,
respectively.
Theoretical lines of (a), (b), and (c)
are calculated by equation~(2) with $C=10^{-4}$ \citep{Tuchman1990}.
The irradiation temperatures of (a), (b), and (c) 
are $T_{\rm irr}$ = 10000~K, 5000~K, and 3000~K, respectively. }
\label{fig10}
\end{figure}

Next, we investigate status of the outer disk (stable or unstable)
assuming that the disk temperature is determined by X-ray heating.
Figure~\ref{fig10} shows the average X-ray luminosity of
the soft (open circles) and hard (filled triangles)
states in the 2--10~keV band 
as a function of the outer radius of the accretion disk.
Here, the outer radius was assumed as 0.35 of 
the binary separation \citep{Smak1982}.
The binary separation was estimated by Kepler's third law
assuming a neutron star mass of 1.4~$\Mo$.
The orbital periods were taken from the
LMXB catalogue \citep{Liu2007} and are listed in table~\ref{tab1}.
The mass of the companion star was taken as 0.4~$\Mo$
although the actual mass is uncertain.
4U~1705$-$44 and XTE~J1709$-$267 were excluded from this analysis
because their orbital periods are not known.

The theoretical lines of (a), (b), and (c) indicate
the ``irradiation disk temperatures'' which depended on X-ray heating
and was calculated by equation (53) of \citet{Tuchman1990}.
The equation (53)
was expressed as 
\begin{eqnarray}
T_{\rm irr} \simeq 10^{3.9}{\rm(K)}
\left( \frac{C}{10^{-4}} \right)^{1/4}
\left( \frac{L}{10^{37}\, {\rm erg\, s^{-1}}} \right)^{1/4}
\left( \frac{R_{\rm d}}{10^{10.5}\, {\rm cm}} \right)^{-1/2}.
\end{eqnarray}
Here, $T_{\rm irr}$ is called the ``irradiation disk temperature,'' 
defined by the irradiation flux as
$F_{\rm irr} = \sigma T_{\rm irr}^4$.
$F_{\rm irr}$
is the irradiation flux on the disk surface from external heat sources. 
The thermal unstable branch (the middle branch of the S-shaped curve)
disappears for strong irradiation with $T_{\rm irr} \geq 10000$~K.
The parameter $C$ denotes the fraction of external heat
caught by the outer disk.
$L$ is the X-ray luminosity of the external heat source
and $R_{\rm d}$ is outer radius of accretion disk.
Mineshige, Tuchman, and Wheeler (1990b) reported that
if the irradiation is a function of the mass accretion rate
and is not so strong ($C \sim 10^{-4}$ and $T_{\rm irr} \leq 5000$~K),
the model can reproduce the observed light variations of dwarf novae
and it is relevant to the outburst of soft X-ray transients
with large amplitudes (roar type).

Thus, we adopted $C = 10^{-4}$ and drew the three theoretical lines
of (a), (b), and (c).
The irradiation temperatures of (a), (b), and (c) 
are $T_{\rm irr}$ = 10000~K, 5000~K, and 3000~K, respectively.
The line of (a) separates the stable and unstable regions.
The left side of the line of (a) indicates the stable region
(fully ionized region).
The region between line (a) and (b) indicate the
unstable region where the normal outburst
(roar type) would not occur
although the borderline depends on the model parameter such as $C$.
Thus, in the region between (a) and (b),
a constant-$\alpha$ disk would be applicable and then
the purr-type instability would occur.
In other words, the observed mini-outbursts
are interpreted as the instability of the purr type in this region.

In figure~\ref{fig10}, except for 4U~1820$-$30,
all four sources have the unstable region between (a) and (b).
It is consistent with the results that mini-outbursts (purr type)
were observed
in 4U~1636$-$536, GS~1826$-$238, and 4U~1608$-$52.
However, in Aql~X-1,
the mini-outbursts were not observed in our observations.
We suggest that it relevant to the short period of hard-high state
defined by Matsuoka and Asai (2013).
For both 4U~1636$-$536 and GS~1826$-$238 (persistent sources),
the outer radii at the hard-state luminosity 
locate on the left side of (b), 
which means the normal outburst (roar type)
would not occur at these luminosities.
It is consistent with the results that 
normal outbursts have not been observed in both sources.
On the other hand, for both 4U~1608$-$52 and Aql~X-1 (transient sources),
the outer radius at the hard-state luminosity 
locate on the right side of (b), where the normal outburst (roar type)
would occur.
It supports that the normal outburst in NS-LMXB
is caused by disk instability in the same way of dwarf novae,
which has been pointed out by \citet{Paradijs1996}.

Variability lasting several tens of days,
which are often observed in the hard state of NS-LMXBs,
might result from a repeat of mini-outburst
although they do not accompany spectral state transitions.
Such variability
was observed in both 4U~1608$-$52 and 4U~1636$-$536.
In 4U~1608$-$52, variabilities without spectral state transitions
were observed during MJD = 56700--56100 (hard-high state).
In 4U~1636$-$536, variabilities in the period after MJD $\sim$ 56000
do not always accompany spectral state transition.
Because the maximum luminosity of the variability
was below the transition luminosity,
inner disk state transition would not occur.

\section{Conclusion}

We investigated the spectral state transitions
in bright NS-LMXBs
using the one-day bin light curves of MAXI/GSC and Swift/BAT.
In four of the sources
(4U~1636$-$536, GS~1826$-$238, 4U~1705$-$44, and 4U~1608$-$52),
we detected small-amplitude X-ray variabilities
with spectral state transitions.
We named these variabilities ``mini-outbursts'' and
interpreted them as disk instability
based on a constant-$\alpha$ disk by Mineshige and Osaki (1985).
In the case of mini-outburst,
the S-shaped curves in the surface density and mass accretion rate diagram,
are modified by the irradiation, and then
the approximation of a constant-$\alpha$ disk would be suitable.
Here, we call mini-outburst as ``purr-type outburst'' 
referring to theoretical studies by Mineshige and Shields (1990a).

We also suggest that variabilities in the hard state
may be a repeat of mini-outbursts
even if spectral state transitions do not occur. 
In this case, the state transition of an inner disk
would not occur because the maximum luminosity of the variability
is below the transition luminosity.

\bigskip

We would like to acknowledge Prof. S. Mineshige for
the useful comment and discussion on this work.
We also would like to acknowledge the MAXI team for MAXI operation
and for analyzing real-time data.
This research was partially supported by JSPS KAKENHI
Grant Number 15H00288.
This research was partially supported by the Ministry of
Education, Culture, Sports, Science and Technology (MEXT),
Grant-in-Aid for Science Research (24340041).


\end{document}